# Topological photonic crystals: physics, designs and applications

Guo-Jing Tang[†], Xin-Tao He[†], Fu-Long Shi, Jian-Wei Liu, Xiao-Dong Chen[*], and Jian-Wen Dong[*]

*School of Physics & State Key Laboratory of Optoelectronic Materials and Technologies, Sun Yat-sen University, Guangzhou 510275, China.*

[†]These authors contributed equally to this work

[*]Corresponding author: chenxd67@mail.sysu.edu.cn, dongjwen@mail.sysu.edu.cn

**The recent research of topological photonics has not only proposed and realized novel topological phenomena such as one-way broadband propagation and robust transport of light, but also designed and fabricated photonic devices with high-performance indexes which are immune to fabrication errors such as defects or disorders. Photonic crystals, which are periodic optical structures with the advantages of good light field confinement and multiple adjusting degrees of freedom, provide a powerful platform to control the flow of light. With the topology defined in the reciprocal space, photonic crystals have been widely used to reveal different topological phases of light and demonstrate topological photonic functionalities. In this review, we present the physics of topological photonic crystals with different dimensions, models and topological phases. The design methods of topological photonic crystals are introduced. Furthermore, we review the applications of topological photonic crystals in passive and active photonics. These researches pave the way of applying topological photonic crystals in practical photonic devices.**

# 1. Introduction

With the development of information technology, the techniques of manipulating and processing light are attracting more and more attentions. There are many interesting phenomena and practical applications such as all-optical switches, optical logic gates, integrated optical circuits. However, some fundamental problems restrict the further development of optical information technology. One of them is the nonnegligible energy loss resulting from fabrication errors of photonic devices. To overcome this problem, people try to develop physical principles besides improving the fabrication techniques. Topological photonics is one of the candidates to solve such a problem. Originating from mathematics, topology distinguishes geometric structures with some quantities that keep invariant during continuous smooth transitions. For example, closed surfaces are classified with the number of "holes" which is characterized by genus. Genus keeps invariant during deformations without tearing or merging the surfaces. By introducing topology into physics, some novel phenomena are robust to perturbations and they have potential of realizing transports which are immune to defects or disorders.

Recently, topology has been widely applied in different fields, including condensed matter [1-6], cold atom [7, 8], acoustics [9-13] and mechanics [14-16]. Topology was introduced to photonics to realize optical analogs of quantum Hall edge states [17] which were firstly observed in electronic systems. Topology and photonics can promote the development of each other. On one hand, compared with electronic systems, photonic systems are convenient to design and tune. So photonic systems are practical platforms to realize novel phenomena of topological physics. On the other hand, topology provides physical principles for improving the performance of photonic devices. Up to now, topological phenomena have been proposed and realized in diverse photonic systems, including photonic crystals (PCs) [16-25], plasmonic systems [18-29], metamaterials [30-42], coupled resonator optical waveguides [43-53], waveguide arrays [54-66], microcavity polaritonic systems [67-74], resonators and waveguide arrays with synthetic dimensions [75-84]. For instance, zigzag chains of plasmonic nanoparticles can support topological edge states and realize the photonic analogue of the Kitaev model of Majorana fermions. The correspondence between the coupled dipole equations in the zigzag plasmonic chain and the Bogoliubov-de-Gennes equations in the quantum wire on top of superconductor was revealed [19]. Chiral hyperbolic metamaterials with topologically protected surface states have been proposed. Such metamaterials possess gapped equifrequency surfaces with nonzero Chern numbers [31]. The topological surface-state arcs in such chiral hyperbolic metamaterials have been observed by the near-field scanning measurement, and confirmed by the backscattering immune propagation of surface waves [35]. Coupled resonator optical waveguides can be used to realize quantum spin Hall effect of light. Hofstadter butterfly and robust edge states were found [43]. The coupled resonator optical waveguide can also be used to realize quantum light sources whose spectrum was less affected by fabrication disorders [51]. With periodic modulations on the direction that light propagates, coupled helical waveguide arrays perform as Floquet topological insulators [54]. Topological valley Hall edge states were also observed in inversion-symmetry-broken honeycomb lattices of waveguides arrays [85]. When magnetic fields are applied, the nontrivial gap were found and photonic quantum Hall effect can be realized in cavity polaritonic systems [69]. The photonic quantum valley Hall effect and quantum anomalous Hall effect were also realized in this system [71]. The proposal of synthetic dimensions provides a method to explore physics in a space with higher dimensions than the realistic structures, such as realizing three-dimensional (3D) Weyl points in the one-dimensional (1D) PCs with the synthetic parameter space [79]. Besides, the synthetic dimension can also be used to construct the 3D topological insulator in a two-dimensional (2D) array of ring resonators [76].

As a significant class of topological systems for controlling the flow of light, PCs play an important role in the realization of topological phases and applications of topological photonics. PCs are optical structures that are periodic in one, two or three dimensions. Due to the Bragg scattering of light in periodic photonic structures, PCs have photonic band structures through which bandgaps can be obtained. Light cannot propagate in the bulk of PCs when its frequency locates in the bandgap. Recently, researchers reveal novel physical principles and solve practical problems by implementing topology in photonics. Since quite a number of quantities of topological physics are defined in the reciprocal space, PCs are considered as good candidate systems to study topological photonics. PCs not only satisfy the requirement of periodicity, but also have the benefits of good light field confinement, multiple adjusting degrees of freedom and ability of integration. Many demonstrations of physical principles and practical applications have been theoretically proposed and experimentally realized in topological photonic crystals (TPCs). In this review, we will introduce the development of TPCs. Physics and designs of 1D, 2D, 3D, and the recent higher-order TPCs are discussed. Especially, we will focus on 2D TPCs which are more favorable in further topological nanophotonics and discuss the quantum Hall, quantum spin Hall and quantum valley Hall effects of light. Lastly, we show the potential applications of TPCs.

## 2. Physics and designs of topological photonic crystals

When designing TPCs, the first and significant step is to choose the topological model [Fig. 1]. Once the model is selected, the dimension and lattice of the PCs are determined, because most models are based on certain dimension and lattice. For example, once the Su-Schrieffer-Heeger (SSH) model is chosen, commonly 1D PCs are considered. For the SSH model, we consider a 1D chain of unit cells consisting of A and B sites. The intra-unit-cell and inter-unit-cell hopping coefficients are $t_1$ and $t_2$, respectively. For a Bloch wave with the wave vector $k$, the Hamiltonian has the form of $H(k) = (t_1 + t_2 \sin ka)\hat{\sigma}_x + (t_2 \cos ka)\hat{\sigma}_y$ where $a$ is the lattice constant, $\hat{\sigma}_i$ are Pauli matrices acting on the subspaces of two states localized on A or B sites. By controlling the hopping coefficients, the SSH chain can be translated between nontrivial ($|t_2| > |t_1|$) and trivial ($|t_2| < |t_1|$) phases. On the other hand, once the Haldane model is chosen, commonly PCs are determined to be 2D and the honeycomb lattice is considered. As shown in Fig. 1, sites at the corners of the unit cell of a honeycomb PC are divided into two sublattices, labeled as A and B. Besides the hopping between nearest neighbor sites, the hopping between second neighbor sites is also considered. The second neighbor hopping contains an additional phase accumulation of $\phi$. And, the on-site energies on A and B sites are $+M$ and $-M$, respectively. For a Bloch wave with the wave vector $\mathbf{k} = (k_x, k_y)$, the Hamiltonian has the form of $H(k) = d_0 \hat{I} + d_1 \hat{\sigma}_x + d_2 \hat{\sigma}_y + d_3 \hat{\sigma}_y$, where $a$ is the lattice constant, $\hat{I}$ is identity matrix, $\hat{\sigma}_i$ are Pauli matrices acting on the subspaces of two states localized on A and B sites,

$d_0 = 2t_2 \cos\phi \left[\sum_i \cos(\mathbf{k} \cdot \mathbf{b}_i)\right]$, $d_1 = t_1 \sum_i \cos(\mathbf{k} \cdot \mathbf{a}_i)$, $d_2 = -t_1 \sum_i \sin(\mathbf{k} \cdot \mathbf{a}_i)$,

$d_3 = M + 2t_2 \sin\phi \left[\sum_i \sin(\mathbf{k} \cdot \mathbf{b}_i)\right]$, $\mathbf{a}_1 = (1,0)a$, $\mathbf{a}_2 = (-1,\sqrt{3})a/2$, $\mathbf{a}_3 = (-1,-\sqrt{3})a/2$,

$\mathbf{b}_1 = (\mathbf{a}_2 - \mathbf{a}_3)/3$, $\mathbf{b}_2 = (\mathbf{a}_3 - \mathbf{a}_1)/3$, $\mathbf{b}_3 = (\mathbf{a}_1 - \mathbf{a}_2)/3$. After the lattice of PC is determined, we should design the unit cells. For example, the unit cell of the 2D square PC can be one dielectric rod in the air background or one air hole in the dielectric background. Last but not least, we need to choose the material according to the operating frequency range. For example, metals can be used as perfect electric conductor in the microwave region, while they are not favor in the infrared region because of the considerable loss. For another example, all-dielectric materials (e.g., ceramic or silicon) are desired for on-chip nanophotonics.

In the following, we will review TPCs in different dimensions, including their effective Hamiltonians and topological invariants, and the theoretical proposal and experimental realizations.

## 2.1. One-dimensional topological photonic crystals

As one of the representative topological models in 1D systems, SSH model can be realized in photonics, and has brought many theoretical and experimental results [86-90]. For example as shown in Fig. 2(a), Meng Xiao *et al.* discussed the photonic SSH model in 1D PCs [90]. A topological invariant, i.e., the Zak phase is defined as [91]:

$$\theta_{Zak} = \int_{BZ} i \int dz \varepsilon(z) u_\mathbf{k}^*(z) \partial_\mathbf{k} u_\mathbf{k}(z) d\mathbf{k}, \qquad (1)$$

where BZ represents the first Brillouin zone, $\varepsilon(z)$ denotes the permittivity distribution and $u_\mathbf{k}(z)$ is the normalized Bloch function. The 1D PC with the inversion symmetry has two inversion centers, and the Zak phase is equal to either 0 or $\pi$ if the origin is chosen to be one of the inversion centers [91]. With the analytical deduction, they showed a rigorous relation that connects the existence of an interface state in the bandgap to the sum of all Zak phases of bulk bands below the bandgap. They constructed two 1D PCs with different Zak phases by adjusting the structural parameters of the unit cell [right panel in Fig. 2(a)]. When two 1D PCs with Zak phases of 0 and $\pi$ are put together, interface states will exist at the interface, which correspond to the transmission peaks within the bandgap. The Zak phase and the resultant interface states in 1D PCs can be borrowed to 2D PCs to achieve multiband waveguides [92]. Two PCs with identical common gaps but different Zak phases were considered and constructed by shifting unit cell of a PC by half the lattice constant. Assembling these two PCs, the multiband waveguide was created [Fig. 2(b)].

Besides the existence of interface states in SSH model, other topological phenomena have been studied in 1D systems. For example, Wei Tan *et al.* established the relation between the topological order and the chirality in metamaterials [93]. The band inversion was experimentally demonstrated, and the interface states were measured [Fig. 2(c)]. This finding provides an example that electromagnetic waves in metamaterials can simulate the topological phenomena in electronic systems. For another example, Alexey A. Gorlach *et al.* designed and tested 1D array of dielectric particles with overlapping electric and magnetic resonances and broken mirror symmetry [94]. The interface states between arrays of meta-atoms with flipped bianisotropy provided the photonic realization of Jakiw-Rebbi modes [Fig. 2(d)]. The local modification of particle bianisotropy results in the modification of coupling constants between neighboring meta-atoms, which offers a promising method to engineer topological states of light. Note that 1D PCs can also simulate the 2D lattice. In 2014, A. V. Poshakinskiy *et al.* discussed topological edge states in 1D PCs with a compound noncentrosymmetric unit cell [95]. With the long-range coupling, this 1D PC was related with an "ancestor" 2D lattice. The time reversal symmetry was broken in the "ancestor" 2D system because the phase accumulations along clockwise and counter-clockwise closed paths were different. The radiative edge states, which decay in time, were shown to survive despite the long-range coupling of resonances and finite lifetime [Fig. 2(e)]. In addition, the Harper model and Kitaev model also have analogue in photonics [2, 96-99]. For example, the quasiperiodic Harper model in optical lattices was experimentally realized [100], and the photonic Kitaev model was studied on the zigzag chain of nanoparticles [19, 101].

Here, we clarify the difference between PCs based on the SSH model and its electronic counterpart. In electronic systems, we focus on states localized at the edges of a SSH chain. Protected by the nontrivial bulk topology and chiral symmetry, the energy of edge state is fixed to be zero even though the coupling coefficient varies. While in PCs based on the SSH model, we focus on states localized at the interface between PCs with Zak phases of 0 and π. The existence of interface state is guaranteed by the rigorous relation between the surface impedance of 1D PC in a gap and the sum of Zak phases of isolated bands below the gap [90], while the frequency of interface state is not fixed because of the absence of chiral symmetry. The frequency of interface state will shift when the structure of PCs (especially the structure of interface) is changed. In this sense, the interface states in PCs based on the SSH model are not as robust as edge states in electronic systems. On the other hand, we note that in some photonic systems such as waveguide arrays [102, 103], coupled microring resonators [104] and coupled PC nanocavity arrays [105], the chiral symmetry can be preserved under certain conditions.

## 2.2. Two-dimensional topological photonic crystals

1D PCs can support interface states and realize novel topological models in photonics. However, to realize robust transport of lights, TPCs with two- or three-dimensional structures should be studied. 2D PCs provide a versatile platform to study topological physics.

### 2.2.1. Quantum Hall PCs

Quantum Hall effect is one of the topological phenomena in condensed matter physics [1, 4]. In quantum Hall effect, current is carried by electrons along the edges of the system, inducing the chiral edge states. Chiral edge states root in the nontrivial topological property of bulk bands. They have a unique directionality and they are robust against disorders [6]. As a seminal work of TPCs, Haldane and Raghu [17, 106] introduced topology into a 2D triangular lattice array of gyroelectric medium. They theoretically constructed analogs of quantum Hall effect in PCs. In a 2D PC, the Chern number of one bulk band is defined as:

$$C_n = \frac{1}{2\pi}\int_{BZ} \Omega(\mathbf{k})d^2k = \frac{1}{2\pi}\int_{BZ} \nabla_\mathbf{k} \times \mathbf{A}^{nn}(\mathbf{k})d^2k, \qquad (2)$$

where $\Omega(\mathbf{k})$ is the Berry curvature, $\mathbf{A}^{nn}(\mathbf{k}) \equiv i\langle u_{n\mathbf{k}}|\nabla_\mathbf{k}|u_{n\mathbf{k}}\rangle$ is the Berry connection, and $u_{n\mathbf{k}}$ is the

periodic part of the Bloch function of eigen modes in the *n*th bulk band. The Chern number characterizes the topological property of photonic bands. The combination of time-reversal and inversion symmetry allows the appearance of Dirac points which are found between the second and third bulk bands of transverse electric (TE) modes [Fig. 3(a)]. The effective Hamiltonian governing states around the Dirac points has the form of a 2D massless Dirac equation:

$$\delta H(\delta \mathbf{k}) = v_D \left( \hat{\sigma}_x \delta k_x + \hat{\sigma}_y \delta k_y \right), \tag{3}$$

where $\delta \mathbf{k}$ measures from the corner of Brillouin zone to the wavevector, $\hat{\sigma}_i$ are the Pauli matrices acting on the subspace corresponding to the two-fold Dirac degeneracy, $v_D$ is the Dirac velocity. Near the Dirac point, the effective Hamiltonian corresponds to a linear dispersion $\omega(\delta \mathbf{k}) - \omega_D = \pm v_D |\delta \mathbf{k}|$ where $\omega_D$ is the Dirac angular frequency. If an external electric field is applied, the permittivity tensor of gyroelectric medium will have nonzero off-diagonal elements. As a result, the time-reversal symmetry is broken and a bandgap is obtained. The effective Hamiltonian will be added a mass term and changes into:

$$\delta H(\delta \mathbf{k}) = v_D \left( \hat{\sigma}_x \delta k_x + \hat{\sigma}_y \delta k_y \right) + \kappa \hat{\sigma}_z, \tag{4}$$

where $\kappa$ is proportional to the difference between the imaginary part of off-diagonal elements in permittivity tensor of gyroelectric medium and that of the background material. The dispersion near the Dirac points becomes $\omega(\delta \mathbf{k}) - \omega_D = \pm \sqrt{v_D^2 |\delta \mathbf{k}|^2 + \kappa^2}$, and it indicates a frequency bandgap with a width of $2|\kappa|$. Below this bandgap, the second bulk band acquires a nonzero Chern number. According to the bulk-edge correspondence, one-way edge states will exist in the bandgap when the boundary is formed between this topologically nontrivial PC (with nonzero Chern number) and a trivial insulator (with zero Chern number).

The proposal of the photonic analogue of quantum Hall effect was first experimentally realized by Wang *et al.* [107, 108]. They showed that one-way modes could be generalized to 2D gyromagnetic PCs without the restriction of Dirac points. A 2D square lattice of PC was designed in gyromagnetic materials [107]. Although Dirac points were not found in this PC, the degeneracy between the second and third transverse magnetic (TM) bands was found at the M point. By applying an external magnetic field, the time-reversal symmetry was broken and the band degeneracy was lifted. A complete bandgap was obtained and it was also characterized by a nonzero Chern number. Reflection-free one-way edge modes were demonstrated at the boundary between this gyromagnetic PC and the trivial perfect electric conductor. To prove their theoretical proposal, they reported the observation of unidirectional backscattering-immune electromagnetic states [108]. The experimental sample involved a 2D gyromagnetic PC consisting of a square lattice of ferrite rods in the air [top panel of Fig. 3(b)]. Chiral edge states were found at the interface between the gyromagnetic PC and the metallic cladding. The robustness of edge states was demonstrated even when a long metallic scatter was placed in the channel [bottom panel of Fig. 3(b)].

Since the theoretical proposal and first experimental demonstration, the quantum Hall PCs have been widely studied in different structures to realize novel light propagation [109-118]. For example, Xianyu Ao *et al.* considered the honeycomb gyromagnetic PCs and showed one-way edge states at the zigzag ribbon boundary [109]. In this honeycomb lattice, the one-way propagation of edge states was insensitive to the imperfections on the zigzag ribbon even if the boundary was exposed to free space. Later then, Yin Poo *et al.* presented an experimental demonstration of such self-guiding electromagnetic edge states [110]. Periodic ferrite rods were arranged in a honeycomb lattice and topological chiral edge states were found in nontrivial bandgap. In addition, multimode one-way waveguides were proposed and realized in gyromagnetic PCs whose bandgaps have large gap Chern numbers (i.e., $|C_{\text{gap}}| > 1$) [111-113].

**2.2.2. Quantum spin Hall PCs**

In the realization of photonic analogue of quantum Hall effect, the large magneto-optical response and external fields are required to break the time reversal symmetry and obtain a topological bandgap. These conditions limit the further applications of quantum Hall PCs, as the magneto-optical effect of gyromagnetic materials is weak in optical frequency and the absorption is nonnegligible. Hence, instead of the time-reversal symmetry breaking TPCs, time reversal invariant TPCs are in need.

As one kind of the time-reversal invariant TPCs, quantum spin Hall PCs were proposed. According to the Kramers theorem, the presence of degenerate partners is necessary for topologically protected $Z_2$ phase in electronic systems. Kramers degeneracy in electronic systems can be formed by the spin-up and spin-down states of the electron. However, it is not the case in photonic systems. The emulation of quantum spin Hall effect (QSHE) in photonics is not straightforward, because photons do not possess a half-integer spin and are not subject to Fermi statistics. Therefore, constructing a pair of pseudospin is an important and the first step when realizing QSHE in photonic systems. Up to date, there are various approaches to construct the photonic pseudospin, such as using electromagnetic dual metamaterials [119-129] or crystalline symmetry of PCs [130-136].

An early work to realize the photonic QSHE in PCs was proposed by Khanikaev *et al.* [119]. When electromagnetic waves propagate in a 2D medium, TM waves with $E_z$ component and TE waves with $H_z$ component have different propagation constants due to different electric permittivity and magnetic permeability tensors (i.e., $\hat{\epsilon}$ and $\hat{\mu}$). However, when electromagnetic dual metamaterials with $\hat{\epsilon} = \hat{\mu}$ are considered, the photonic pseudospin states can be constructed by the linear combinations of TM and TE waves. These two pseudospin states are doubly degenerate and are time-reversal partners. When a triangular PC with four-fold degenerate Dirac points is considered [Fig. 4(a)], its effective Hamiltonian has the form of:

$$\delta H(\delta \mathbf{k}) = v_D \left( \hat{\tau}_z \hat{s}_0 \hat{\sigma}_x \delta k_x + \hat{\tau}_0 \hat{s}_0 \hat{\sigma}_y \delta k_y \right) + \zeta \hat{\tau}_z \hat{s}_z \hat{\sigma}_z, \tag{5}$$

where $\hat{\sigma}_i$, $\hat{s}_i$ and $\hat{\tau}_i$ are Pauli matrices acting on the subspaces of two-fold Dirac degeneracy, pseudospin and valley. The last term opens a bandgap with a width of $2|\zeta|$. The spin Chern number, the quantized topological invariant charactering quantum spin Hall PCs, is defined as $C_s = (C_+ - C_-)/2$, where $C_+$ ($C_-$) is the Chern number of pseudospin-up (pseudospin-down) states and can be derived as $C_\pm = \pm \text{sgn}(\zeta)$. Then, the circular rod within the unit cell is filled with split ring resonators which have the bianisotropic response [inset of Fig. 4(a)]. Specifically, the bianisotropic response of this metamaterials is described by the constitutive relations: $\bm{D} = \hat{\epsilon}\bm{E} + i\hat{\chi}\bm{H}$ and $\bm{B} = \hat{\mu}\bm{H} - i\hat{\chi}^T\bm{E}$ with the bianisotropic tensor $\hat{\chi}$. With this nonzero bianisotropy, the four-fold degenerate Dirac points will be opened and a photonic topological insulator (PTI) with a nonzero spin Chern number is finally obtained. This work makes 2D bianisotropic PCs a powerful platform for studying the fundamental physics of time-reversal invariant topological states.

After the bianisotropy-induced quantum spin Hall PTI was proposed, it was experimentally realized by Wen-Jie Chen *et al*. [120]. Instead of utilizing bianisotropy which was intrinsically weak in metamaterials, they introduced a large 'effective' bianisotropy in a uniaxial metacrystal waveguide [Fig. 4(b)]. Within the metacrystal waveguide, the waveguide modes are naturally coupled. It leads to an effective bianisotropic response which is related to the order of waveguide modes and the thickness of waveguide. In the experimental sample, they designed two kinds of non-resonant and ε/µ-matching meta-atoms, arranged them in the hexagonal lattice, and designed a PTI and a photonic ordinary insulator (POI). By forming an edge between PTI and POI, the spin-polarized one-way edge states were observed. To test the robustness of edge states, they introduced defects in the waveguide and found that high transmittances of both $E_z$ and $H_z$ fields were maintained. At the same time, Tzuhsuan Ma *et al*. proposed another photonic structure, i.e., bianisotropic metawaveguides to achieve bianisotropy-induced quantum spin Hall PTIs [121]. The metawaveguide consisted of a parallel-plate metal waveguide which was filled with an array of metallic cylinders. These embedded metallic cylinders were connected to the top or bottom metal plates. The finite gap between the rods and the top/bottom plates gave the required bianisotropy. In addition, using two topologically nontrivial metawaveguides, a topological channel which supported spin-locked edge states was constructed. These spin-locked edge states could be guided without reflections along sharp bends. The experimental realization of such metawaveguides was carried by Xiaojun Cheng *et al.* [125]. The robust and reconfigurable transmission of spin-locked edge states was confirmed. In addition, they also demonstrated a pseudospin selective wave divider showing the potential application of the spin-locked surface modes. Very recently, an all-dielectric metasurface emulating QSHE has been implemented [137]. By using dielectric materials, the limitation caused by Ohmic losses in the earlier design based on metallic metamaterials can be overcome. To emulate the spin degree of freedom, double degeneracy between electric and magnetic dipolar modes were achieved by carefully tuning the structural and constituent parameter of metesurfaces. They broke the vertical symmetry by introducing a circular notch on one of the flat faces of the cylinders, which essentially causes the spin-orbit interaction of light and makes the metasurface into a PTI. The robust edge states revealing the spin-locked properties of this PTI was visualized by the near-field spectroscopy.

Most quantum spin Hall TPCs introduced above need metallic materials, which limits their application in higher frequency because of the loss of electromagnetic waves in metal is considerable. The all-dielectric design is needed for the practical applications of on-chip devices. To achieve this, Long-Hua Wu and Xiao Hu demonstrated that all-dielectric PCs with crystalline symmetry could also emulate the QSHE [130]. They deformed a honeycomb lattice of cylinders into a triangular lattice of unit cells with six hexagon-arranged cylinders [top panel of Fig. 4(c)]. With this deformed supercell, the Dirac cones at the K/K' points are folded to doubly degenerate Dirac cones at the Γ point. In this $C_{6v}$ symmetric PCs, the role of pseudospins is played by the positive and negative angular momenta of the wave function. By solving the Maxwell equations, degenerate orbital-like *p*- and *d*-wave shaped eigen-fields were found. By shrinking or expanding the hexagonal cluster of cylinders, doubly degenerate Dirac cones were opened and two topologically distinct bandgaps were obtained [bottom panel of Fig. 4(c)]. The evolution of band structure and the associated topological phase transition can be captured by the effective Hamiltonian:

$$\delta H(\delta \mathbf{k}) = v_D \left( \hat{s}_z \hat{\sigma}_x \delta k_x + \hat{s}_0 \hat{\sigma}_y \delta k_y \right) + \left( M + Bk^2 \right) \hat{s}_0 \hat{\sigma}_z, \qquad (6)$$

which has a similar form with the Hamiltonian of Bernevig-Hughes-Zhang model. The spin Chern number charactering the pseudospin-up and pseudospin-down can be evaluated as $C_\pm = \pm[\text{sgn}(M) + \text{sgn}(-B)]/2$. Since the parameter *B* is typically negative, the spin Chern number will be nonzero when the sign of parameter *M* is positive.

The above crystalline PCs has a simple design which makes it a promising platform for photonic applications. Later then, Yuting Yang *et al*. implemented this kind of spin Hall PCs by the periodic $Al_2O_3$ cylinders, as shown in Fig 4(d), in the microwave region [131]. Similarly, doubly degenerate Dirac points were designed at the Γ point by the deformation from the honeycomb lattice to a triangular lattice. A sharp bending interface between a TPC and a trivial one was constructed and the propagation of the pseudospin-polarized electromagnetic states was visualized. In addition, a square-shaped four-antenna array was designed to selectively excite one of the two pseudospin polarized states. Although the $C_{6v}$ symmetric lattice using dielectric cylinders is more friendly to optical devices for its simple structure compared with the metamaterial-based PCs, it has some difficulties to address when going to applications. For example, metallic mirrors (which are undesirable for photonic devices) are used to have a good confinement of electromagnetic waves along the out-of-plane direction. To design TPCs with good out-of-plane confinement but without metallic components, Sabyasachi Barik *et al*. presented a free-standing silicon PC slab with a bandgap for TE-like modes [132]. With the similar symmetry consideration, the honeycomb lattice of the PC slab with triangular air holes is also deformed into the triangular lattice. Later in 2018, they demonstrated the topological properties, such as spin-locked transmission and robust transmission [136]. This nanostructured silicon slab constitutes a platform to explore many-body quantum physics with topological protection. Recently, the operating wavelength of spin Hall TPCs have reached visible spectral range. Siying Peng *et al*. experimentally fabricated a hexagonal photonic lattice of silicon Mie resonators, as shown in Fig. 4(e) [133]. By mapping the local optical density with a deeply subwavelength resolution, bulk bands in visible spectral range were measured. This work presents the feasibility to realize practical silicon-based topological geometries that are commensurate with planar silicon-based integrated photonic technology.

### 2.2.3. Quantum valley Hall PCs

Quantum spin Hall PCs are time-reversal invariant systems and are favor in application in optical frequency. Recently, another degree of freedom, "valley", has been exploited and valley photonic crystals (VPCs) were proposed and widely studied [27, 127, 138-152]. In 2016, Tzuhsuan Ma and Gennady Shvets proposed an all-Si valley Hall PTI [143]. When the inversion symmetry is broken, a complete bandgap appears in TE bulk bands [Fig. 5(a)]. The effective Hamiltonian has the form of:

$$\delta H(\delta \mathbf{k}) = v_D \left( \hat{\tau}_z \hat{\sigma}_x \delta k_x + \hat{\sigma}_y \delta k_y \right) + \lambda \hat{\tau}_z \hat{\sigma}_z, \qquad (7)$$

where $\hat{\sigma}_i$ and $\hat{\tau}_i$ are Pauli matrices acting on the subspaces of two-fold Dirac degeneracy and valley.

The topological index $C_{K/K'}$ is defined as the integration of Berry curvature in the half Brillouin zone close to K/K' and the valley Chern number $C_v = C_K - C_{K'}$. The sign of $\lambda$ determines the valley Chern number by $C_v = \text{sgn}(\lambda)$ and the resultant nonzero valley Chern number indicates the nontrivial topology. The zigzag interface between two VPCs with opposite valley Chern numbers supports edge states. Robust delay lines were designed by using the scattering immunity property of edge states. When edge states are refracted at the termination between PCs and the air, different effects of zigzag and armchair terminations were illustrated.

Valley-contrasting physics in PCs was also discussed by Xiao-Dong Chen *et al.* [144]. The TM waves in all-dielectric honeycomb PCs were considered, and the opposite phase vortexes of bulk states at the *K* and *K'* valleys were revealed. The valley chiral bulk states could be selectively excited by controlling the chirality of sources. The valley contrasting Berry curvature distribution was calculated and was found to be localized around K and K' points. Valley-dependent edge states and their broadband robust transport were illustrated by the backscattering suppression at the Z-shaped bend. Jian-Wen Dong *et al.* designed VPCs with valley-dependent spin-split bulk bands [139]. The inversion symmetry was broken by making two rods in a unit cell have opposite bianisotropy. The photonic valley Hall effect of was observed, even though the Chern number, spin Chern number and valley Chern number are zero. The valley spin locking behavior was represented by splitting of flows of opposite spins along the ΓK and ΓK' directions [Fig. 5(b)]. Because of the valley-spin locking behavior, the net spin flow could be selectively excited by choosing LCP or RCP source. TPCs with broken inversion symmetry and time reversal symmetry have also been studied [153-157].

With the increasing theoretical studies, experimental realizations of VPCs become essential [27, 145-148]. Compared with experiments in infrared and even visible regions, microwave experiments are more convenient since the samples are simpler to fabricate. For example, Xiaoxiao Wu *et al.* directly observed the valley-polarized edge states in designer surface plasmon crystals [27]. When the mirror symmetry is broken [Fig. 5(c)], this designer surface plasmon crystal becomes a valley-Hall PTI and supports topological edge states. A beam splitter was used to selectively excite valley-polarized edge states, whose amplitude ratio spectrum between each port showed the valley-polarization property. As another example, Zhen Gao *et al.* proposed a VPC that supports surface electromagnetic waves and demonstrated such VPC on a single metal substrate [145]. Arranged as a graphene-like honeycomb lattice, the unit cell of this VPC consisted of two metallic rods [Fig. 5(d)]. This platform can be used to study bulk and edge valley transport whose experimental observations were measured by the near-field mapping. The bulk states were excited by an incident narrow beam, and the valley-locked beam splitting was observed.

Topological photonics is expected to be applied in on-chip integrated photonics. With the exploration of silicon-on-insulator nanofabrication technology, VPCs working in infrared region have been experimentally realized [158-161]. Mikhail I. Shalaev *et al.* fabricated an optical topological insulator that realized the QVH effect [158]. The mirror symmetry was broken by changing the size of two triangular holes in the unit cell [Fig. 5(e)]. Confirmed by the transmittance spectrum of straight and trapezoidal channel, topologically protected propagation of edge states was realized. Xin-Tao He *et al.* realized topological transport in a silicon-on-insulator VPC slab which has a TE-like bandgap at the telecommunication wavelength [159]. The VPC slab supported photonic edge states that localized within the plane of slab [Fig. 5(f)]. The phase vortex of magnetic fields and ellipticity angle of electric fields in VPCs were simulated. The robust transport of edge states was realized in flat, Z-shaped, and Ω-shaped interfaces, confirmed by both simulated and experimental transmission spectra. In addition, the unidirectional coupling of valley-dependent edge states can be excited by a chiral source. Based on this property, photonic routing was realized by using a microdisk to excite phase vortex and couple to different interfaces. On the other hand, terahertz technology attracts more and more attention for its potential applications in the next generation of communication technologies. Yihao Yang *et al*. realized the robust transport of terahertz waves in VPCs based on the all-silicon chip [Fig. 5(g)]. They demonstrated the error-free communication which enables the real-time transmission of uncompressed 4K video through a highly twisted domain wall [160].

## 2.3. Bound states in the continuum in photonic crystals

Within the aforementioned 1D and 2D TPCs, light propagates along the in-plane directions. However, for structures with a finite height (e.g., gratings and PC slabs), light may leak out of the systems along the out-of-plane radiation. Nontrivial topology of TPCs protects the light propagation from losses in the in-plane direction, but not in the out-of-plane direction. To inhibit out-of-plane radiation, the concept of bound states in the continuum (BIC) is introduced into photonics. BIC was first proposed in quantum mechanics [162] and can also be realized in classical wave systems. There are different kinds of BICs, classified by their physical origins and constructing methods, including BICs from symmetry or separability, from parameter tuning in resonance systems, and from inverse construction [163]. BICs and quasi-BICs have been found in photonic systems such as PC slabs [164-202], gratings [203-218], metamaterials [219-226], coupled waveguide arrays [227-229], array of spheres or rods [230-233]. Here, we will mainly introduce the topology of BICs in PCs.

Bo Zhen *et al*. demonstrated that the symmetry protected and accidentally occurring BICs in PC slabs were at the vortex centers of the polarization direction of far field radiation [170]. Taking the z axis as the vertical direction of the PC slabs, electric fields of resonance modes have the form of $\mathbf{E_k}(\rho,z) = e^{i\mathbf{k}\cdot\rho}\mathbf{u_k}(\rho,z)$, where $e^{i\mathbf{k}\cdot\rho}$ is the propagation term, $\rho = x\hat{x} + y\hat{y}$ is the in-plane spatial coordinate and $\mathbf{u_k}(\rho,z)$ is the periodic part of wave functions. We consider the projection of polarization of $\mathbf{u_k}$ as a polarization vector $\mathbf{c}(\mathbf{k}) = c_x(\mathbf{k})\hat{x} + c_y(\mathbf{k})\hat{y}$ [Fig.6 (a)]. The BICs will exist at the vortex centers of $\mathbf{c}(\mathbf{k})$ in the *k* space. We consider the angle of polarization vector $\phi(\mathbf{k}) = \arg[c_x(\mathbf{k}) + ic_y(\mathbf{k})]$, then the topological charge carried by a BIC can be defined as the winding number of the polarization vectors, namely $q = \frac{1}{2\pi}\oint_C d\mathbf{k} \cdot \nabla_\mathbf{k}\phi(\mathbf{k})$, where *C* is a closed path in the *k* space that goes around the BIC in an anticlockwise direction. The right column of Fig. 6(a) shows the polarization vector distribution around BICs carrying the topological charge of +1 (top) and -1 (bottom). The topological charge is quantized and conserved, which makes it possible to predict and understand the behaviors of BICs such as bouncing, annihilation and generation of topological charge.

Besides 2D PC slabs, 1D periodic arrays of dielectric spheres can also support BICs [231, 232]. E. N. Bulgakov *et al.* analyzed the BICs in periodic arrays of dielectric spheres [top of Fig. 6(b)] and demonstrated that BICs are related to the phase singularities of the quasimode coupling strength [232]. As shown in the bottom of Fig. 6(b), BICs exist at the vortex center of coupling strength's phase. The nodal lines, which represent zero real part (red dash line) and zero image part (blue line), cross at the vortex center. At this singularity point, the coupling strength is zero, i.e. BICs exist. BICs are robust against any parameter variations preserving the periodicity and rotational symmetry. However, when the symmetry is broken, the BICs will degrade into quasi-BICs. K. Koshelev *et al.* proposed a theory for asymmetric periodic structures and demonstrated the connection between the BICs and Fano resonances [222]. Based on the explicit expansion of the Green's function of open systems into eigenmode contributions, the analytical approach describing light scattering by arrays of meta-atoms was revealed. The relation between the conventional Fano formula and the reflection (and transmission) coefficients was also demonstrated.

The experimental characterization of the topology of BICs has also been demonstrated [185-192, 217-220]. For example, BICs in gratings were experimentally presented with polarization vortex by H. M. Doeleman *et al.* [234]. They designed and manufactured a 1D $Si_3N_4$ grating supporting a BIC and confirmed its existence with reflection measurements. They measured the polarization state of reflection around the BIC with k-space polarimetry and the expected vortex of polarization was observed [bottom of Fig. 6(c)]. The polarization vortices in momentum space of 2D plasmonic crystals were observed by Y. Zhang *et al.* [185]. They mapped out the detail information of radiative states. The momentum-space vortices of polarization were identified by the winding patterns in the polarization-resolved isofrequency contours and diverging quality factors. BICs in 1D and 2D periodic structures are promising to suppress the out-of-plane radiation. J. Jin *et al.* theoretically proposed and experimentally demonstrated the out-of-plane-scattering losses suppression in PC slabs by using BICs [188]. The quality factor in the experiment reached 12 times higher than those obtained with standard designs. Besides the out-of-plane radiation suppression, BICs in gratings can also be utilized to realize unidirectional guided resonances [218]. Such resonances emerged when a pair of half-integer topological charges bounce into each other [Fig. 6(d)]. The positions of topological charges in wave vector space could be controlled by changing the tilted angle of sidewalls.

## 2.4. Three-dimensional topological photonic crystals

In 3D TPCs, one of the most significant concepts is the Weyl point which is a monopole of Berry flux. As the 3D extension of the 2D Dirac point, a Weyl point is a degenerate point where two linear dispersion bands intersect in 3D momentum space. However, 2D Dirac cones can be easily gapped by breaking either the parity inversion symmetry (*P*) or the time-reversal symmetry (*T*), while 3D Weyl points are protected against any perturbations. Therefore, there are various intriguing properties of Weyl points, which attracts tremendous attention [38, 39, 235-245]. The effective Hamiltonian of Weyl points has the general form of

$$\delta H(\delta \mathbf{k}) = \sum_{i,j} v_{ij} \hat{\sigma}_i \delta k_j, \qquad (8)$$

where *i*, *j* = *x*, *y*, *z*. The topological property of Weyl point is characterized by its chirality χ = sgn[det($v_{ij}$)]. The topological property is also characterized by the Chern number *C* of a closed surface enclosing a single Weyl point, which has the relation with the chirality as $C = -\chi$.

In 2013, Ling Lu *et al.* proposed a photonic realization of Weyl points by using a 3D perturbed double-gyroid PC whose bulk bands are shown in Fig. 7(a) [237]. The bulk bands of the unperturbed double-gyroid PC contained a frequency-isolated threefold degeneracy among the third, fourth and fifth bands at the Γ point. By replacing a part of the gyroid material with two air-spheres, this threefold degeneracy was lifted without breaking *P* or *T* symmetry. Furthermore, by breaking the *P* or *T* symmetry, Weyl points can be obtained. When undergoing pure *P*-breaking perturbation (i.e., place only one air-sphere on one of the gyroid), two pairs of Weyl points with different chirality will emerge. Based on the theoretical proposal, they presented the experimental observation of Weyl points [236]. In the experiment, the angle-resolved transmission measurement was performed to probe the dispersions of the 3D bulk states. As a result, 3D bulk states projected along the [101] direction and the Weyl point along the Γ-H direction were mapped out experimentally. In addition to using the gyroid structure, other approaches were also explored to realize Weyl points in 3D PCs. One of them is to use stacked 2D slabs with designed interlayer coupling, which is compatible with planar fabrication technology. In 2016, Wen-Jie Chen *et al.* designed and fabricated a 3D PC possessing single and multiple Weyl points [238]. To obtain 3D Weyl points, they started with the 2D honeycomb lattice with Dirac cones, then stacked these honeycomb lattices periodically in the z-direction and introduced chiral interlayer coupling [Fig. 7(b)]. By doing this, Weyl points with topological charges of ±1 were found at the K/K' point while Weyl points with topological charges of ±2 were found at the Γ point. In 2018, Biao Yang *et al.* presented a PC of saddle-shaped metallic coils, serving as an ideal Weyl system where the Weyl points exist at the same energy and are well-separated [38]. By the near-field measurements, the helicoidal structure of the topological surface states were observed. As shown in the right panel of Fig 7(c), the equifrequency contour shows the presence of four symmetrically displaced elliptical bulk states with the same size located along the diagonal directions. There are other means to construct photonic Weyl points such as creating the Weyl points in a synthetic 3D space [76, 79, 246].

Due to the linear crossing bulk bands, 3D Weyl PCs are gapless photonic materials. On the other hand, 3D gapped TPCs with topological bandgaps have also been studied. In 2016, Alexey Slobozhanyuk *et al.* discussed a 3D all-dielectric topological photonic metacrystal which was the photonic analog of 'weak' TIs [247]. With carefully designed structure, the electromagnetic duality between electric and magnetic fields was ensured, and the photonic pseudospin could be constructed. Then by breaking the z-mirror symmetry, the effective spin-orbit interaction was introduced [Fig. 7(d)]. As a result, the photonic metacrystal experienced a topological transition and a complete 3D photonic bandgap was obtained. Without the lossy metal-based components, this all-dielectric design serves as a promising platform for optical applications. With the idea of constructing a 'weak' TI by stacking the 2D quantum spin Hall insulators, Yihao Yang *et al.* presented the experimental realization of a 3D TPC with metallic patterns on printed circuit boards [248]. In their experimental sample, a 3D array of metallic split-ring resonators was designed and the resonance-enhanced bianisotropy gave a strong photonic spin-orbit coupling [Fig. 7(e)]. Therefore, the width of the topological bandgap was greater than 25%, which was much greater than many bandgap widths in previous 2D or 3D TPCs. Utilizing two dipole antennas inserted inside the sample, the transmission spectra and field distribution along the domain wall were measured. Searching on 3D TPCs may pave the way for future applications in topological photonic cavities, circuits, lasers in 3D structures.

## 2.5. Photonic crystals with higher-order topology

The above discussed TPCs host robust edge states predicted by the bulk-edge correspondence principle: A $d$-dimensional TPC hosts ($d$-1)-dimensional edge states. Recently, higher-order topology has been introduced in electronic insulators and brought into classical wave systems, featuring with in-gap corner/hinge states which that do not obey the usual bulk-edge correspondence principle [249-260]. Such higher-order topology has been realized in photonic [261-283], acoustic [284-291], and electric circuit systems [292-294].

Photonic systems with higher-order topology have been proposed and realized in honeycomb [261, 275], square [263-268, 276-280] and kagome lattices [262, 269]. The 2D SSH model can be utilized to describe the 2D second-order PCs with 0D corner states [251, 264, 267, 268, 295]. Let us first consider the square lattice of PCs whose unit cells consisting of four sites [Fig. 8(a)]. By changing the distance between two neighboring sites in same and adjoining unit cells, the intra-unit-cell and inter-unit-cell hopping coefficients (i.e. $t_a$ and $t_b$) can be adjusted. Similar to Zak phase in 1D systems, we can define 2D Zak phase:

$$Z_j = \int dk_x dk_y \text{Tr}\left[\hat{A}_j(k_x, k_y)\right], \quad (9)$$

where $j = x, y$, $\hat{A}_j(k_x, k_y) = i\langle u(\mathbf{k})|\partial_{k_j}|u(\mathbf{k})\rangle$ and $|u(\mathbf{k})\rangle$ is the periodic Bloch function. When the mirror symmetry is preserved along the $x$ and $y$ directions, $Z_x$ and $Z_y$ are quantized to be 0 or $\pi$. So there are four topological phases which are characterized by $\mathbf{Z} = (0, 0), (0, \pi), (\pi, 0)$ and $(\pi, \pi)$. Corner states can be found at the corners of the bulk sample of phase characterized by $\mathbf{Z} = (\pi, \pi)$ surrounded with the sample of phase characterized by $\mathbf{Z} = (0, 0)$. Such corner states have been experimentally realized in PCs operating at microwave region [267, 268] and infrared region [266, 279]. In 2D PCs with square lattice, two configurations were used to construct higher-order topological insulating phases [267]. Different from systems with negative coupling between sites in previous studies, the dielectric PCs shown in Fig. 8(b) have no negative coupling. Both topological edge states and corner states were observed with near-field scanning technique. Topological corner states can also be supported in PC slabs [268]. Different from TM and TE modes in 2D PCs, PC slabs have more complex polarizations. A perfect electric conductor was used to filter out TE-like modes and a complete bandgap was found [Fig. 8(c)]. PC slabs with four different 2D Zak phases were designed. In the square sample constructed by PC slabs with Zak phases of (0, 0) and $(\pi, \pi)$, topological corner states were observed. Corner states were also realized in TPCs operating at infrared region. Based on corner states, a PC nanocavity with the resonance peak at 1079 nm was designed and fabricated [266]. The presence of topological corner states could be predicted with bulk-edge and edge-corner correspondences [Fig. 8(d)]. The corner mode was tightly localized with a quality factor over 2000. Quantum dots could be used to excite topological corner states in PC nanocavities. It was reported that the coupling between quantum dots and topological corner states would affect the emission properties of quantum dots [279]. The photoluminescence intensity and emission rate would be enhanced when the quantum dot was on resonance with the corner state.

In addition to the 2D SSH model, we can also derive the polarizations which characterize the higher-order topology with Wannier representation [249, 250, 276, 282]. The 0D corner states can also be found in 2D quantum spin Hall PCs [275]. Recently, a new class of topological corner states originating from long-range interactions are found in PCs with breathing kagome lattice [269]. The tight binding model, which only considers the nearest-neighbour coupling and ignores the interactions of longer distance, can predict the existence of corner states, which well confined at the corner of topologically nontrivial samples surrounded by the trivial ones, so called type I corner states. However, there is another kind of corner states, so called type II corner states which have field profiles similar to that of edge states but tend to localized to corners when the coupling detuning parameter increases [Fig. 8(e)]. To analyze type II corner states, the next-nearest-neighbour coupling needs to be considered.

Here, we clarify the robustness of corner states in PCs with higher-order topology. Because of the absence of chiral symmetry, corner states in PCs based on 2D SSH model are not as robust as their counterparts in electronic systems because the frequencies of corner states in PCs will change with the variation of structural parameters. The Q factor will also vary with the structural modification [296]. However, being protected by certain symmetries, the high-order topological corner states in PCs have considerable robustness. For example, in PCs with $C_6$ symmetric breathing honeycomb lattice, the high-order topological states are protected by lattice symmetries and have certain degree of robustness against disorders and long-range interactions which break the chiral symmetry [270]. In PCs with breathing kagome lattice, the frequencies for type I corner states are insensitive to the variation of structural parameters because of the protection of generalized chiral symmetry. While type II corner states are not pinned at a fixed frequency [269].

## 2.6. Topological photonic crystals with synthetic dimension, deformation and modulation

In addition to the real spatial dimensions, topological phenomena can also appear in TPCs with synthetic dimensions [84]. With synthetic dimensions such as frequency and structural parameters, researchers can use PCs to realize topological phenomena in higher dimensions. For example, Qiang Wang *et al*. constructed 1D interface states protected by the synthetic 3D Weyl point [79]. The 1D PCs had unit cells constructed with four alternate layers of $HfO_2$ and $SiO_2$ [Fig. 9(a)]. The generalized Weyl points in optical frequency region were realized in the parameter space spanned by the Bloch wave vector $k$ and two structural parameters $p$ and $q$. An effective Weyl Hamiltonian for the parameters around conical degeneration point was derived. Besides, optical interface states were found between PCs with synthetic Weyl points and reflection substrates. The reflection phase of truncated PC was experimentally measured and exhibited vortex around a synthetic Weyl point in the parameter space. The existence of interface states is guaranteed by the reflection phase vortex property of synthetic Weyl points. Researchers can even realize topological phenomena in 4D parameter space. By using synthetic dimension, a topological one-way fiber was designed and realized based on a 3D magnetic Weyl PC with double gyroids [245]. A 3D Chern crystal was obtained by adding plain modulation of volume fraction. This 3D Chern crystal is characterized by three first Chern numbers defined on three momentum planes respectively. Then, by changing the plane modulation into helical modulation, such structure supports modes with backscattering immunity and works as a one-way fiber. The topological invariant of such one-way fiber is the second Chern number which is defined in 4D parameter space spanned by three wave vectors, $k_x$, $k_y$, $k_z$, and the angle of modulation $\theta$. Another kind of modulation, Kekulé modulation, was also studied in PC fibers [297]. To begin with, researchers added an extra air hole in a silica PC fiber with triangular lattice of air holes, and this gaps two nodal lines into Weyl points in *k* space. Then, for the sake of gapping the entire nodal line and keeping the Weyl degeneracies, supercell modulations of perturbing the lattice with generalized Kekulé pattern [298, 299] with a vortex phase were operated. Such modulations coupled two nodal lines and annihilated them into a bandgap. Then the existence of mid-gap defect modes was guaranteed to propagate at the core of such Dirac vortex fiber. And the number of guiding modes is just the winding number of vortex modulation. Recently, Kekulé modulation was used to construct Dirac vortex cavity [300].

The Dirac vortex is constructed with the angular modulation of the structural deformation. The deformations arranged in other forms were also studied. Some deformations can add a gauge field in the effective Hamiltonian of PCs. When the deformations cause a $\delta\mathbf{k}$ shift of band structure, that is, operate the transform of $\mathbf{k} \rightarrow \mathbf{k} + \delta\mathbf{k}$, a gauge field vector potential $\mathbf{A} = \delta\mathbf{k}$ will be added. The gauge field vector potential will induce an artificial effective magnetic field of $\mathbf{B} = \nabla \times \mathbf{A}$. For example, F. Deng *et al.* proposed and realized the control of valley pseudospin currents with a gauge field added by straining the graphene structure of PCs [301]. The linear distortion along the *y* direction resulted in constant effective magnetic fields with opposite signs at K and K' valleys. The valley-dependent propagations in bended baths were observed. Another example is the chiral zero mode in inhomogeneous 3D Weyl metamaterials which was designed by rotating the saddle-shaped metallic coils at different positions with different angles [39]. The rotation of metallic coils caused the rotation of Weyl points with a same angle around the $k_z$ axis. With the gradual change of rotation angle along the *x* direction, an artificial magnetic field along the z axis was added and it had opposite directions for Weyl points at $k_x > 0$ and $k_x < 0$. The presence of gauge field was experimentally confirmed and the zero-order chiral Landau level with one-way propagation was observed.

The effective magnetic fields can be generated by not only spatial modulation but also temporal modulation. K. Fang *et al.* studied a resonator lattice with harmonically temporal modulated coupling constants between resonators [75]. When the distribution of the modulation phases was tuned as shown in the left panel of Fig. 9(f), an effective gauge potential corresponding to a uniform effective magnetic field was introduced. The effective magnetic field resulted in a Lorentz force for photons and the presence of topologically protected one-way edge states. The robust transportation could exist without using magneto-optical effects.

# 3. Photonic applications

To further explore the potential applications of TPCs, many functionalities and prototypes of passive or active photonic devices have been presented, based on the novel properties of TPCs. In this section, we will give a brief introduction of several representative potential applications of TPCs, such as robust waveguide, lasing and other active functionalities.

## 3.1. Robustness of topological photonic crystals

Robustness is one of the most important characteristics of TPCs. For example, due to the robustness of edge states of 2D TPCs, electromagnetic waves can smoothly propagate along the domain wall between two topologically distinct PCs, even when disorders, defects or sharp corners exist. In the quantum Hall PCs under the external electric or magnetic fields, one-way chiral edge modes can be achieved [17, 107-110, 116, 302]. As shown in Fig. 10(a), these edge states are unidirectional and are robust against obstacles. Even when a metallic scatterer is placed in the interface, the left-ward propagating edge modes do not suffer backscattering. Similar to chiral edge states in quantum Hall PCs, the helical edge states can be found in quantum spin Hall PCs. Note that such helical edge states have pseudospin-dependent robustness against certain defects that do not induce the coupling between two pseudospins [119-124, 130-133]. For example in Fig. 10(b), interfaces of electromagnetic-dual topological metacrystals supported helical edge states that exhibit spin-polarized one-way propagation of light, which is robust against defects with some missing rods [119]. Later, the robustness of valley-dependent edge states in quantum valley Hall PCs were also discussed [143-145, 158-160, 303, 304]. Xin-Tao He *et al.* showed valley-dependent edge states in a silicon-on-insulator VPC slab [159]. Robust transport of light with the telecommunication wavelength was observed along Z-shaped bends, showing the flat-top high transmission of ~10% bandwidth [Fig. 10(c)]. In addition to 2D systems, 3D TPCs have also been explored by observing the robust transport of surface states [237-239, 247, 248]. For instance, Yihao Yang *et al.* realized a 3D TPC and mapped out both the gapped bulk band structure and the Dirac-like dispersion of the photonic surface states [248]. Then they demonstrated robust propagation of surface waves along a non-planar surface. Robust surface states were found in the bulk topological bandgap and flew robustly along the surface, including around the two sharp corners [Fig. 10(d)].

Here, we discuss the robustness of edge states in different kinds of 2D TPCs. The transport of edge states in quantum Hall PCs is robust against most defects because of the broken time reversal symmetry. In the TPCs with preserved time reversal symmetry, the situations are complex. The robust properties of edge states in quantum spin Hall PCs based on electromagnetic dual metamaterials and crystalline symmetry are different. For quantum spin Hall PCs based on electromagnetic dual metamaterials, the transports of edge states are robust against various defects, including the sharp bending of interface, cavity obstacle and disordered domain in adjacent PCs [119]. However, for quantum spin Hall PCs based on crystalline symmetry, the robust transports were only reported at the sharp bends of 60°, 90° and 120° [130-132, 136]. In VPCs, the robust transports of edge states are usually demonstrated in waveguides with 60° or 120° bends [138, 148, 160]. However, even though edge states in VPCs can only transport robustly at 60° and 120° bends, one can also construct interfaces with arbitrary turning angles with series of 60° and 120° bends and perform robust transport in such interfaces [160].

We should note that backscattering-immune transport is just one of the representative phenomena to demonstrate the robustness of TPCs. There are some other phenomena related to topological robustness, such as the near-perfect out-coupling [140, 143] and the robustness of corner states against structural parameters [249, 250, 261, 262, 265-270]. When the inter-valley scattering is suppressed in quantum valley Hall PCs, the perfect coupling of edge states into homogeneous medium can be achieved [143]. This is used to show topologically protected refraction of spin-valley-locked kink states [140]. When the valley pseudospin is conserved, the kink states exhibit nearly perfect out-coupling efficiency into directional beams [Fig. 10(e)]. For PCs with higher-order topology, corner states which are robust against defects can be found [261, 262, 265, 269, 270]. In 2020, Mengyao Li *et al.* reported robust corner states in higher-order topological PCs with kagome lattice that exhibits non-zero bulk polarization [269]. Protected by the generalized chiral symmetry, type I corner states are pinned to zero energy even when some structural parameters are changed [Fig. 10(f)].

## 3.2. Passive functionalities of topological photonic crystals

In TPCs, based on their enormous potential in light manipulation, many applications have been proposed and expected to realize high performance optical devices.

Optical cavity, as one of the important elements of photonics, is widely used in many areas of physics and engineering, including coherent electron–photon interactions, low-threshold lasers, nonlinear optics. To design high-performance cavity in photonic integrated circuits, many researches have focused on all-dielectric PC slabs through finely engineering point defects in a precise location of the PC slab. By using such defect modes of PCs, researchers have proposed high-Q nanocavities with optical wavelength sizes [305]. The introduction of topology into the PCs brought a new mechanism to design optical cavities. Using robust edge modes in TPCs, researchers proposed cavities in which lights travel along the edge [109, 134, 306], and edge modes will not be influenced by defects at the pathway of light. Besides, several topological cavities based on real-space topology [307], Zak phase [308-310], Jackiw-Rebbi model [94], Dirac-vortex topology [300], and the recent higher-order topology [266, 279] have also been proposed. For example, Dirac-vortex topological cavities were designed and realized by applying the generalized Kekulé modulations on the silicon PC slabs [300]. The generalized Kekulé modulations were introduced by adding a displacement on three sites of the unit cell with $C_3$ symmetry. Figure 11(a) depicts the distribution of magnitude and directional angle charactering the displacement in a Dirac-vortex cavity. With such modulation, the Dirac-vortex cavity was confirmed to have scalable mode areas, arbitrary mode degeneracies, vector-beam vertical emission and compatibility with high-index substrates. The free spectral range was demonstrated to be large, which is important for the stabler single-mode operation, a higher spontaneous emission factor and a wider spectral tuning range.

Optical filters play an important role in communication systems. For example, a narrowband flat-top reflection filter is in need for achieving the wavelength sensitivity in wavelength-division multiplexing systems [311]. An all-pass transmission filter is useful for applications such as optical delay or dispersion compensation [312]. Combining waveguide modes and cavity modes in 2D PCs, researchers can achieve narrow band filters [313]. Taking advantage of the unidirectionality and robustness of edge states in TPCs, Jin-Xin Fu *et al.* constructed a unidirectional band-stop filter and a unidirectional channel-drop filter that could selectively process an optical signal propagating only along a particular direction [115]. In the designed four port filter [Fig. 11(b)], the transmission spectra showed that most of the energy of the waveguide mode transmitted from port 1 to port 3 at the resonance frequency of 14.72 GHz. In comparison, away from the cavity resonance frequency, almost all waves went to port 2. There were hardly any waves passing through port 4, as the waves toward port 4 of the lower waveguide were forbidden. Besides, valley filters based on valley-dependent edge states were also proposed [71, 127].

Beam splitters are optical devices that split one beam of light into two beams. It is a crucial part of many optical measurement systems such as interferometers, and also finds widespread applications in the fiber optic communications. By designing the waveguide topography or interface, people have presented several splitters made by PCs [314-318]. In 2014, Scott A. Skirlo *et al.* designed a continuously tunable power splitter based on quantum Hall phases in PCs with large Chern numbers [111]. As shown in Fig. 11(c), they demonstrated the continuously tunable power splitter implemented with $C_{gap} = 2$ and $C_{gap} = 1$ gyromagnetic PCs bordered on the top by a metallic wall. Since one-way modes could not be backscattered, they only scattered into each other, changing their amplitudes and phases. The height of the metal scatterer controlled the total mode profile at the junction, and consequently the power splitting between channels 1 and 2. The total power efficiency of the splitter was always 100%. Besides, both pseudospin-polarized and valley-polarized power splitters have also been achieved in quantum spin Hall [319, 320] and valley Hall [139, 144, 145, 147] PCs.

A circulator is a passive, non-reciprocal three- or four-port device, in which the signal entering any port is transmitted to the next port in rotation. It can be achieved in 2D magnetic PCs [321, 322]. The key point of this type of circulator is the coupling between a point defect and the waveguide channels, results in narrowband performance. TPC brings a new way to realize circulating of electromagnetic waves. For example, Tzuhsuan Ma *et al.* proposed a four-port broadband circulator based on nonreciprocal topological edge waves between QSH-PTIs and a QH-PTI [138]. They constructed the sample by embedding a QH-PTI island inside an interface-containing sea of QSH-PTIs [Fig. 11(d)]. Combining different kinds of domain walls which support different one-way edge states, the unique pathway could only guide incident light from each port to the next port. This kind of circulator has the advantages of broadband frequency and robustness.

In the optical communications, wavelength-division multiplexing is a technology which multiplexes a number of optical carrier signals onto a single optical fiber/waveguide by using different wavelengths. A wavelength-division multiplexing system uses a multiplexer at the transmitter to join several signals together and a demultiplexer at the receiver to split them apart. Wavelength-division multiplexer and demultiplexer can be realized in PCs by carefully designing the coupling between waveguides and cavities [323-326]. Recently, Guo-Jing Tang *et al.* proposed a frequency range dependent photonic detouring based on VPCs with dual bandgaps [327], which had potential applications in wavelength division multiplexers. The designed VPCs have dual bandgaps which are located at two different frequency ranges. Frequency range dependent edge states in two bandgaps were found in the domain walls between different VPCs. The channel between VPC1 and VPC2 supported edge states in both bandgaps. The channel between VPC3 and VPC2 only supported edges states in the first bandgap while the channel between VPC1 and VPC3 only supported edge states in the second bandgap. Therefore, As shown in Fig. 11(e), lights of edge states locating in different bandgaps traveled along different paths and spatially separated.

TPCs can also be used to manipulate the flow of light in the out-of-plane direction. Based on the topological properties of BICs, B. Wang *et al.* proposed and realized an approach to generating optical vortices, which had spiral wavefronts and screw phase dislocations [192]. Using the topological vortex-like response of polarization in momentum space centered at BICs of PC slabs, they induced Pancharatnam-Berry phases and spin-orbit interaction in light beams. Such vortex generators operating in momentum space has almost homogeneous structures without a real-space center to align at the incident beam center.

### 3.3. Lasing in topological photonic crystals

In addition to the above introduced passive functionalities of TPCs, there are active functionalities such as lasing behaviors. Recently, lasing in topological photonic structures has been theoretically analyzed and experimentally realized in PCs [105, 328-335], coupled ring resonators [104, 336-339] and polariton micropillars [74, 89, 340]. Laser in different topological systems has different advantages such as single-mode lasing, high efficiency and immunity to disorders and so on. Here, we focus on the lasing in TPCs.

**3.3.1. lasing in cavities based on 1D states in TPCs**

Lasing in cavities based on the 1D edge states in quantum Hall [328], quantum spin Hall [337] and quantum valley Hall [331] topological systems has been realized. In quantum Hall TPCs, the nonreciprocal lasing can be exhibited [328]. As shown in Fig. 12(a), the PC was made of InGaAsP quantum wells bonded on yttrium iron garnet, which was grown on gadolinium gallium garnet. The time-reversal symmetry was broken by the yttrium-iron–garnet substrate under an external magnetic field. The PC on the inner side of the cavity had a square lattice and was topologically nontrivial. The outer side PC had a triangular lattice and was topologically trivial. A defect waveguide was coupled to the cavity. Resulting from the QH effect, the topological edge states in this structure had properties of unidirectional propagation and backscattering immunity from imperfections and sharp corners. Thus, the cavity could be shaped arbitrarily. Edge states with only one propagating direction were supported in the cavity when the external magnetic field is fixed. The unidirectional lasing with an isolation ratio of 11.3 dB was achieved when the optical pump power density is 0.9 $\mu W/\mu m^2$. The topological insulator laser in quantum spin Hall system was realized in coupled ring-resonator array systems [337]. The coupled ring-resonator array was fabricated with InGaAsP and the emission intensity profiles are shown in the inset of the right of Fig. 12(b). The topological properties gave rise to single-mode lasing, robustness against defects and considerably higher slope efficiencies. As shown in the left panel of Fig. 12(b), the slope efficiency of topological array was higher than that of its trivial counterparts and the enhancement reached about threefold in experiment. In the emission spectra of topological and trivial array [right of Fig. 12(b)], the single-mode character could be observed with narrower linewidths compared with trivial array. Furthermore, with S-chiral microresonators assembled, the unidirectional lasing without magnetic fields was enforced. The topological lasers introduced above need optical pumping. However, in many practical applications, such as quantum cascade lasers, electrical pumping is required. Recently, the electrically pumped topological laser based on valley edge modes has been realized [331]. The VPC was fabricated on the active medium of a THz quantum cascade laser wafer. The PC had a triangular lattice of quasi-hexagonal holes that break inversion symmetry. The lasing cavity was constructed by a triangular loop of topological waveguide. The lasing spectrum had robust regularly spaced emission peaks that persist under certain disturbances. For example, when an external waveguide was coupled to the cavity and worked as a directional outcoupler, the spectra of topological modes coupled to the left and right sides had same peaks, compared with the spectra of non-topological modes which are completely different [Fig. 12(c)]. Topological insulator laser in telecommunication region with quantum valley Hall PCs have also been proposed [341] and realized [342].

Compared with conventional lasers based on ring resonators, most lasers with cavities based on 1D states in TPCs have robustness against certain defects, that is, the defects have less influence on the lasing effect. Besides, quantum Hall and quantum spin Hall systems can support single-mode lasers. The single-mode property will prevent the mode competing and, as a result, improve the stability and efficiency of lasers.

### 3.3.2. lasing in cavities based on 0D states in TPCs

Lasing can also be realized in cavities based on 0D states in TPCs. For example, a nanocavity was formed at the interface between two PCs with distinct Zak phases [334]. The nanocavity had a high Q factor and a near-diffraction-limited mode volume. With gain from semiconductor quantum dot, the lasing effect with the single-mode property and a high spontaneous emission coupling factor of about 0.03 had been observed. The spontaneous emission coupling factor was orders of magnitude larger than those of conventional semiconductor lasers, which comes from the strongly optical confinement. The unit cell of the PC contained two rectangular air holes in GaAs slab with a width of 1.6$a$ and a thickness of 0.64$a$, where $a$ is the lattice constant. These two air holes had a width of 0.5$a$ and lengths of $d_1$ and $d_2$, where $d_1 + d_2 = 0.5a$ is conserved. The inversion symmetry was preserved, which resulted in the quantization of Zak phase to either 0 or $\pi$. By changing $d_1$ and $d_2$, the Q factor and mode volume could be adjusted [Fig. 13(a)]. The robust single-mode lasing is beneficial to develop single-mode lasers using broadband gain media. Lasing can be also exhibited by topological edge states in 1D L3 nanocavity array embedded in a 2D PC [105]. The PC was composed of an InAsP/InP multiple-quantum-well epilayer. The chain of nanocavity could be described by SSH model. The coupling strengths were controlled by the distance between adjacent cavities. The structure with a winding number of one supported topological edge states. The lasing of edge states was demonstrated and the modal robustness was observed. In theory, the energy stored in the cavity remained close to 1 when structural fluctuations were introduced. In experiment, the single-mode lasing was preserved with a minor drift of lasing frequency.

Recently, with the exploration in higher order topology, corner states in 2D PC have been proposed and realized. Corner states are localized at the corner of interface between two topologically distinct PCs. Such localization property can be exploited in the realization of nanocavity for topological lasers [272, 332]. Based on the second-order corner state, low-threshold topological nanolasers have been realized [332]. The corner state was induced at the corner of interface between TPCs with 2D Zak phases of ($\pi$, $\pi$) and (0, 0). Lasing behavior of corner state was observed at 4.2K. The topological laser had a low threshold of approximate 1 μW and a high spontaneous emission coupling factor of approximate 0.25 [Fig. 13(b)]. Although the Q factor and resonance frequency of corner state were sensitive to disorder around the corner, the corner state existed even with harsh perturbations in the bulk of PC.

Compared with conventional lasers based on nanocavities, lasers with cavities based on 0D states in TPCs can survive under perturbations that do not change the topological properties of bulks, even though the resonance frequencies and Q factors are susceptible to variations near the cavities. Besides, if the chiral symmetry is approximatively preserved in some conditions, the resonance frequencies have certain stability to the variation of structural parameters [105, 334].

### 3.3.3. vertical emission lasing in TPCs

In addition to the lasing confined in the in-plane of TPCs, vertical emission lasing in TPCs was also explored. The lasing in a pumped BIC cavity has been realized [176]. As shown in Fig. 13(c), The PC was constructed with InGaAsP multiple quantum wells cylindrical nanoresonator array. By changing the radius of nanoresonators, the quality factor of cavity modes approached infinity to form a BIC when the radius was 528.4 nm. The lasing wavelength of the BIC cavities scaled with the radii of the nanoresonators corresponding to the theoretical prediction for the BIC mode. As the lasing in BIC cavity has been realized, the control of such lasing behavior is in need for further applications in dynamic manipulation. By employing BICs, researchers could also realize the ultrafast control of vortex micorlasers [191]. Perovskite-based vortex microlasers with application in ultrafast all-optical switching at room temperature were realized. The vortex beam lasing could be switched to linearly polarized beam lasing, or vice versa, by breaking the fourfold rotational symmetry with changing the spatial deviation of pumping region or using two-beam configuration [left of Fig. 13(e)]. A parameter K was introduced as the ratio of linear polarization to characterize the switching time. As shown in the right of Fig. 13(e), the switching time was 1 to 1.5 picoseconds. Besides, the energy consumption was orders of magnitude lower than previously demonstrated all-optical switching.

In addition to BICs, the cavity based on band inversion of quantum spin Hall PCs can also be used to generate vertical emission laser [330]. The topological bulk laser cavity consisted of topological nontrivial PC on the inner side and the trivial one on the outer side. At the cavity boundaries, reflection was induced by the band inversion. Because the band inversion only occurs around the Γ point, the lasing mode was selected and the lasing emission is directional. The lasing threshold was as small as ~4.5 kW cm$^{-2}$ according to the integrated output power as a function of pump intensity [Fig. 13(d)]. The divergence angle was less than 6°. The topological bulk laser worked at room temperature in a single mode. The side-mode suppression was over 36 dB and reached the requirement of most practical applications. The cavity size, threshold and linewidth also reached the requirements in most applications.

### 3.4. Active functionalities of topological photonic crystals

In photonics, both gain and loss will make the systems be non-Hermitian. Note that loss is unavoidable in various photonic devices. Therefore, the study of non-Hermitian topological photonic systems is quite important. Based on the exploration of non-Hermitian systems [343-345], efforts were taken to realize topological photonic systems with non-Hermiticity [346-350]. In addition to topological lasing (with gain media), there are many researches of non-Hermitian photonic systems focusing on TPCs [351-354]. TPCs with gain and loss were also realized experimentally. For example, 1D TPC based on the SSH model, was constructed in dielectric microwave resonators [86]. A microwave antenna was placed at the interface between two configurations of SSH model. Zero mode was observed at the interface and could be enhanced by adding absorption at proper sites in two topologically distinct configurations. The robustness of such midgap state was exhibited by the similar intensity profiles under structural disorders.

The topological corner states in non-Hermitian PCs were also studied recently. As shown in Fig. 14(a), gain and loss were introduced in PCs whose superlattice contains four cylinders arranged in square [354]. Two kinds of structures, one was diagonal non-Hermiticity and the other was parallel non-Hermiticity, were demonstrated. Topological corner states could be presented in such non-Hermitian PCs which contained a nontrivial region enclosed by a trivial region. Four near-degenerate midgap corner states emerged in both diagonal and parallel non-Hermitian structures. For diagonal non-Hermiticity system, there were two of the four corner states experiencing thresholdless phase transition as the non-Hermitian coefficient γ grows. The rest two states would be nondegenerate when γ continues to grow and reach an exceptional point. After reaching the exceptional point, the real parts of all four eigenfrequencies degenerated. For parallel non-Hermiticity system, before γ reached the exceptional point, four eigenfrequencies were real and the system was Hermitian. As γ continued to grow and reached the exceptional point, the eigenfrequencies transited into two complex conjugate pairs and the system turned into non-Hermitian.

In principle, material absorption and radiation are two basic mechanisms for the origin of loss in photonic systems. The radiation loss is common in PC slabs. The study of radiation loss can help to clarify its effects on the modes in PC slabs. H. Zhou *et al*. explored the non-Hermitian topological properties of exceptional points [355]. Consider a 2D-periodic PC slab with a finite thickness, where resonance modes experience radiation loss, the Dirac point would split into a pair of exceptional points. The pair of exceptional points were connected with a bulk Fermi arc, the open-ended contour of degeneration of the real part of eigenfrequencies [Fig. 14(b)]. The half-integer topological charges of polarization were discovered in the far-field radiation around the bulk Fermi arc. These topological charges described the direction and number of times the polarization vector winds around a point or line singularity.

In addition to the unavoidable loss in photonic systems, sometimes we will advisedly increase the loss to actively control the optical signals. An optical switch is a device that selectively switches optical signals from one channel to another in the time domain. It has been achieved in PCs based on nonlinear optics [356, 357], self-imaging principle [358] and dynamic shift of the photonic bandgap [359]. In TPCs, robust edge states are able to wrap around sharp corners, which allows optical signals pass through a bended channel. Otherwise, if the operation frequency of signal locates at the range of bulk states, it will scatter into the bulk. Using this property, researchers have designed optical switches based on tunable or reconfigurable TPCs [28, 360-362]. Figure 14(c) shows an optical switch made by reconfigurable TPC [360]. The dynamic control of topological edge states was demonstrated by applying the external electric field to modify the refractive index of a liquid crystal background medium. It resulted in a change of the spectral position of the photonic bandgap and the topological edge states. Thus, the optical switch could be realized by controlling external electric field. When external electric field was applied, i.e. ON-state, there was no bulk state allowed for both topologically nontrivial and trivial regions, so light was efficiently guided along the rhombus-shaped path. When the external electric field was absent, i.e. OFF-state, bulk states were present at the operation wavelength, enabling light scattering into the bulk.

Researchers can also control the transmission of optical signal with another beam of light. The optically tunable TPC has been proposed and realized [362]. The all-optical free-carrier excitation, which allowed for fast refractive index modulation, was utilized to dynamically control the transmission in a silicon-based TPC. When the ultra-violet pump beam illuminated, there are two mechanisms of the free carrier excitation, i.e. single-photon absorption and two-photon absorption, leading to the change of refractive index [Fig. 14(d)]. Resulting from the changes of real and imaginary parts of the refractive index, a blue shift up to 20nm occurred in transmission spectrum and the transmittance was reduced by approximately 85%. Such all-optical control mechanism could achieve switching times of the order of nanoseconds.

The refractive index in Ref. [362] is controlled by free carrier excitation. Other refractive index modulation methods can also be explored, such as Kerr nonlinearity, which has a nonlinear response. Nonlinear optics is of great importance in frequency operation, optical information processing, and quantum optics. Many topological photonic systems have been proposed and applied in nonlinear optics [51, 273, 363-369] . For instance, third-harmonic generation has been used in the characterization of TPCs [369]. To overcome the limitations in the experimental characterization of topological photonic structures, nonlinear topological photonics could be employed by observing nonlinear light conversion in a topological photonic nanostructure. By measuring the generated third-harmonic light under pump beams with different frequencies and polarizations, the selective imaging of bulk states and edge states were demonstrated [Fig. 14(e)]. The proposed nonlinear imaging, compared with linear approach, had benefits of superior contrast, sensitivity, and large imaging area. Similar nonlinear process was also demonstrated with TPCs supporting topological corner states [273]. Another example is the frequency mixing process of one-way edge states in TPCs [370]. The band structure and topological properties of two-dimensional PCs with hexagonal symmetry were analyzed in detail. Topological protected one-way edge states within bandgaps at different frequencies were exhibited. By tailoring the edge configuration, phase matching of nonlinear optical processes was achieved. Frequency processes, such as second-harmonic generation and third-harmonic generation, were demonstrated. Besides, other behaviors such as slow-light effects and counter-propagating mode interaction were also exhibited in this setup.

In addition to the interaction between photons in the nonlinear optics, the interaction between photon and electron in polaritonics can also be utilized in active functionalities of TPCs. The strong coupling between edge states in TPCs and excitons in monolayer $WS_2$ has been demonstrated by W. J. Liu *et al.* [371]. The topological polaritonic system was formed by monolayer $WS_2$ coupled with quantum spin Hall PCs with breathing honeycomb lattice. The topological polaritons in this system can be observed without deep cryogenic temperature and external magnetic fields. The spin momentum-locked helical property of the topological polaritons was demonstrated by real-space image and angle-resolved spectral measurement. More recently, Li *et al.* has demonstrated the strong coupling of bulk photonic states and excitons in 2D materials, which leads to topological transition and formation of topological polaritonic phases characterized by non-zero spin Chern numbers [372]. In addition to topological photonic edge states in optical frequencies, topological photonic states in midinfrared frequency range were also studies by S. Guddala *et al*. Midinfrared topological photonic modes were confirmed to have ability of coupling with the phononic modes in hexagonal boron nitride [373]. Besides, with the photoluminescence of quantum spin Hall PCs coupled with monolayer $WSe_2$, the valley polarization of edge polaritons was confirmed. The monolayer of 2D transition metal dichalcogenides have exhibited nonreciprocity under circularly polarized pumping [374], which have potential applications in tunable all-optical devices. Considering the tunability and active property of 2D transition metal dichalcogenides, the possibility of integrating 2D materials with existing topological photonic systems provides a novel and attractive direction in topological polaritonics and active topological photonic applications.

Recently, topological photonics has been applied in quantum optics to generate and control topologically protected quantum states [102, 136, 273, 375-379]. Topologically robust electromagnetic modes were realized in a 2D array of ring resonators. Their vacuum fluctuations were utilized to create a quantum light source [51]. The spectrum of generated photons was much less affected by disorder. The correlated photon pairs were generated by spontaneous four-wave mixing. Compared with their topologically trivial 1D counterparts, such correlated photon pairs outperformed in terms of spectral robustness. Another application of TPCs in quantum optics is the interface between single quantum emitters and topological photonic states [136]. A quantum emitter was efficiently coupled to topological edge states at the interface between two topologically distinct all-dielectric PCs, as shown in Fig. 14(f). The power was reduced below the quantum dot saturation power to isolate a single quantum emitter within the topological edge state. The chiral nature of the topological edge states was demonstrated. When an external magnetic field was applied in the out-of-plane direction, Zeeman splitting would be induced in the quantum dot emitter under an ultralow temperature. The excited states split into two nondegenerate states with opposite circular polarizations. These two opposite circular-polarized states would propagate in different directions along the waveguide. Besides, the topological edge states had robustness to bends. According to the results of second-order correlation measurement, the routed photons are confirmed to be single photons.

## 4. Conclusions and outlook

In summary, we have reviewed the fundamental physics, optical design and promising applications of TPCs. The studies of TPCs experience three stages: the theoretical proposal, the experimental observation, and the functionality demonstration. Along this roadmap, we firstly introduce the topological physics in 1D, 2D, 3D TPCs, BICs in PC slabs, PCs with higher-order topology, and TPCs with synthetic dimension, deformation and modulation. Then, we represent the potential applications of TPCs in passive devices such as waveguides, cavities, circulators, filters, splitters, wavelength-division multiplexing prototypes, vortex generators, and in active devices such as lasing, non-Hermitian functionalities, applications in non-linear optics and quantum optics.

PC provides a convenient platform for realizing those intriguing phenomena of topological physics experimentally. For example, non-Hermitian topological photonic systems are attracting more and more interests of researchers [380-389]. The combination of non-Hermitian and topological physics inspires the observation of intriguing phenomena and practical applications. For another example, the real-space lattice defects in topological systems have also inspired the interests of researchers recently [390-395]. The edge states and bound states have been observed in PCs with bulk disclinations [391]. The disclination states with fractional spectral charge have been realized in PCs [392, 395]. The researches of classical wave systems with lattice defects will promote the study of phenomena that interact topology in momentum space and real space. Recently, the bilayer photonic crystals have attracted many researchers' attention [396-403]. Bilyer photonic crystals provide a platform to study the novel physical properties in bilayer systems with Moire superlattices.

Based on the theoretical exploration, more and more researches devote the topological physics to photonic device prototype, particular for on-chip phenomena demonstration in optical frequency region. The all-dielectric design of topological photonic system which is compatible with CMOS technics will be hopefully applied in practical photonic devices. At the same time, some ineluctable challenges should be overcome when dealing with the issues in integrated optics, such as insertion loss and out-of-plane radiation. Because of the mismatching between the modes in TPC waveguides and conventional rectangular or ridge waveguides, the loss at the interface between two kinds of waveguides is significant. On the other hand, because of the limitation on materials utilized in available technology, some functional demonstrations that require propagation modes in TPC slabs have to use edge states above the light cone, which will introduce out-of-plane radiation and increase the loss. The insertion loss and propagation loss are unevadable problems in the practicality process of TPCs. In addition to robust transport, there are more and more properties of light manipulation realized in TPCs. Making the best of topology induced property in particular functionality is quite important. Compared with topologically trivial systems, the advantage of topological systems, such as the ability of reaching extreme parameters, should be explored. Comparing topological systems with conventional devices in practical operating conditions is necessary for the application of topological photonics in practicable devices. Another challenge for the practical applications, especially the on-chip applications, of TPCs is the dynamic regulation and reconfiguration with technologies compatible with existing devices. Compared with regulation based on optical induced effects, electrical control is more common in practical devices. Recently, there have been some TPCs with electric controlled reconfigurability theoretically proposed and experimentally realized in microwave region [360, 361, 404]. For the application of topological photonic systems in optoelectronic technologies, exploring the topological physics and topological-related light manipulation in optoelectronic materials and devices is also necessary [405-407]. Further exploration of the dynamic tunable and reconfigurable TPCs operating at infrared region will promote the applications of topological photonic systems in optoelectronic devices, such as modulators, photodetectors and electric pumped lasers.

## Availability of data

The data that support the findings of this study are available from the corresponding author upon reasonable request.

## Disclosure statement

No potential conflict of interest was reported by the authors

## Funding

This work was supported by National Natural Science Foundation of China (Grant Nos. 62035016, 12074443, 61775243), Guangdong Basic and Applied Basic Research Foundation (Grant No. 2019B151502036), Natural Science Foundation of Guangdong Province (Grant No. 2018B030308005), and Fundamental Research Funds for the Central Universities (Grant Nos. 20lgjc05, 20lgzd29).

**Figures**

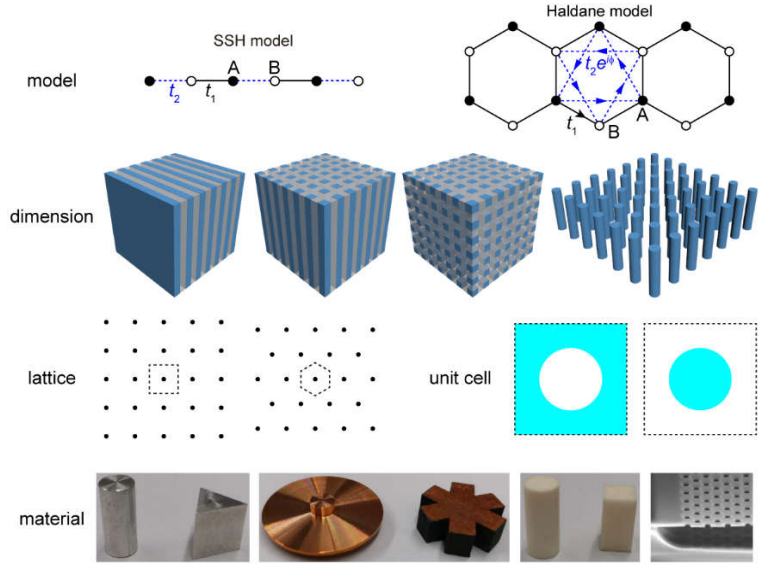

**Figure 1. The designing procedure of TPCs.** To achieve a TPC, one should first choose the model with nontrivial topology, and then determine the dimension, lattice, unit cell, and the constitutive materials of PCs.

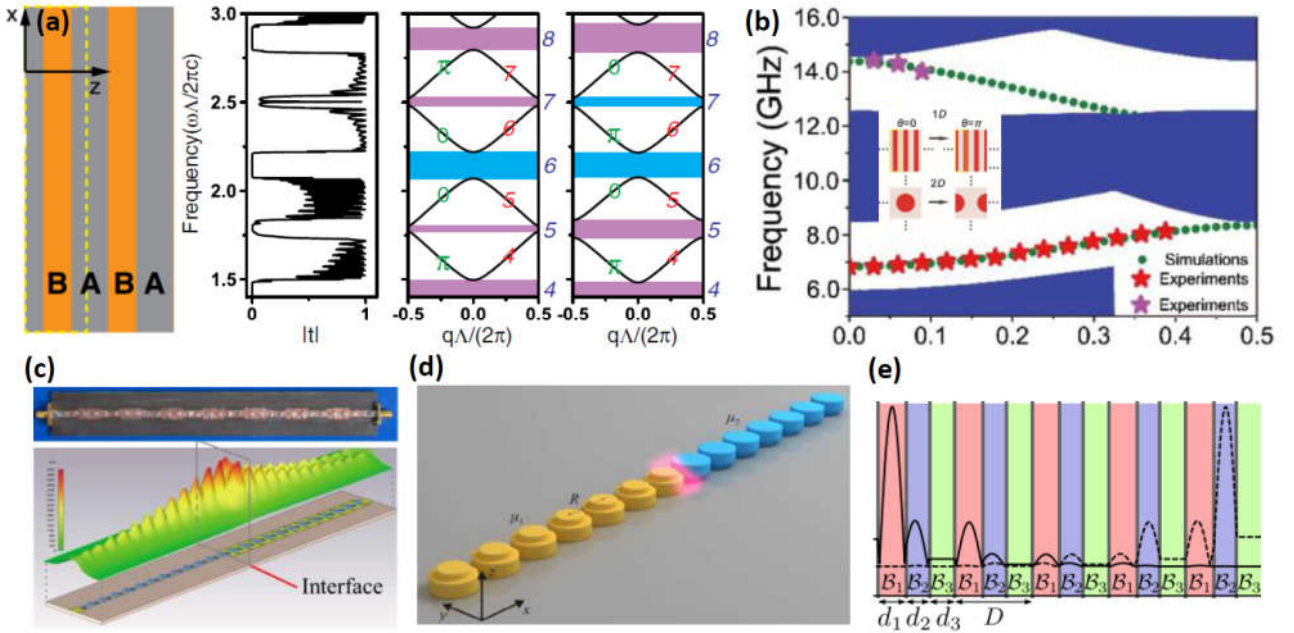

**Figure 2. One-dimensional TPCs.** (a) Inversion symmetric 1D PC and the transmission spectrum of the structure between PC1 and PC2 whose bulk bands and the associated Zak phase are given [90]. (b) Multi edge states in photonic domain wall between two 2D PCs with different Zak phases [92]. (c, d) Observation of interface states in the (c) 1D metamaterials [93] and (d) 1D dielectric rods with opposite bianisotropy [94]. (e) Location of boundary states in the 1D PCs [95].

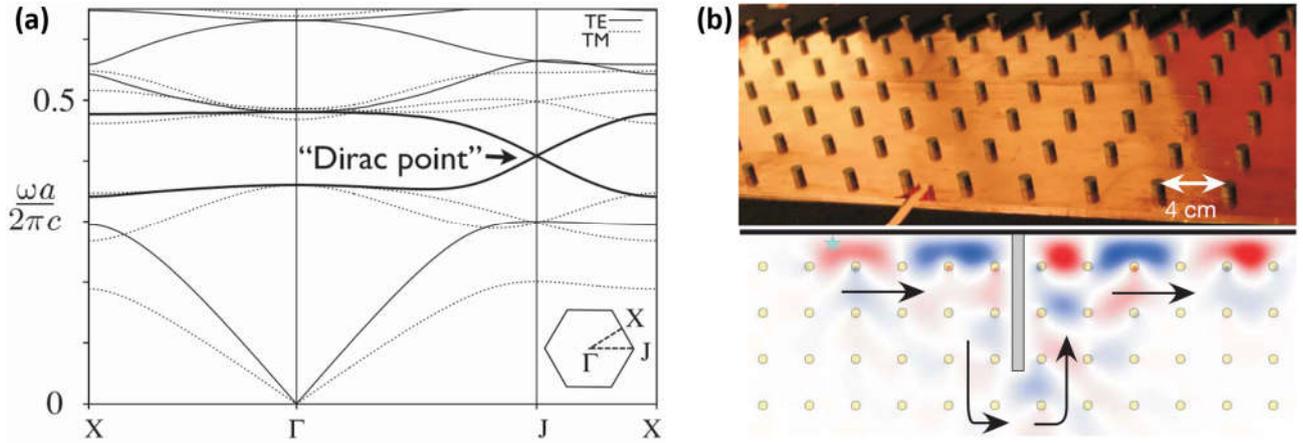

**Figure 3. Quantum Hall PCs.** (a) The photonic bands of a 2D triangular lattice of gyroelectric PC [17]. (b) The waveguide using in the microwave experiment and the robust transmission of edge states against an inserting perfect electric conductor [108].

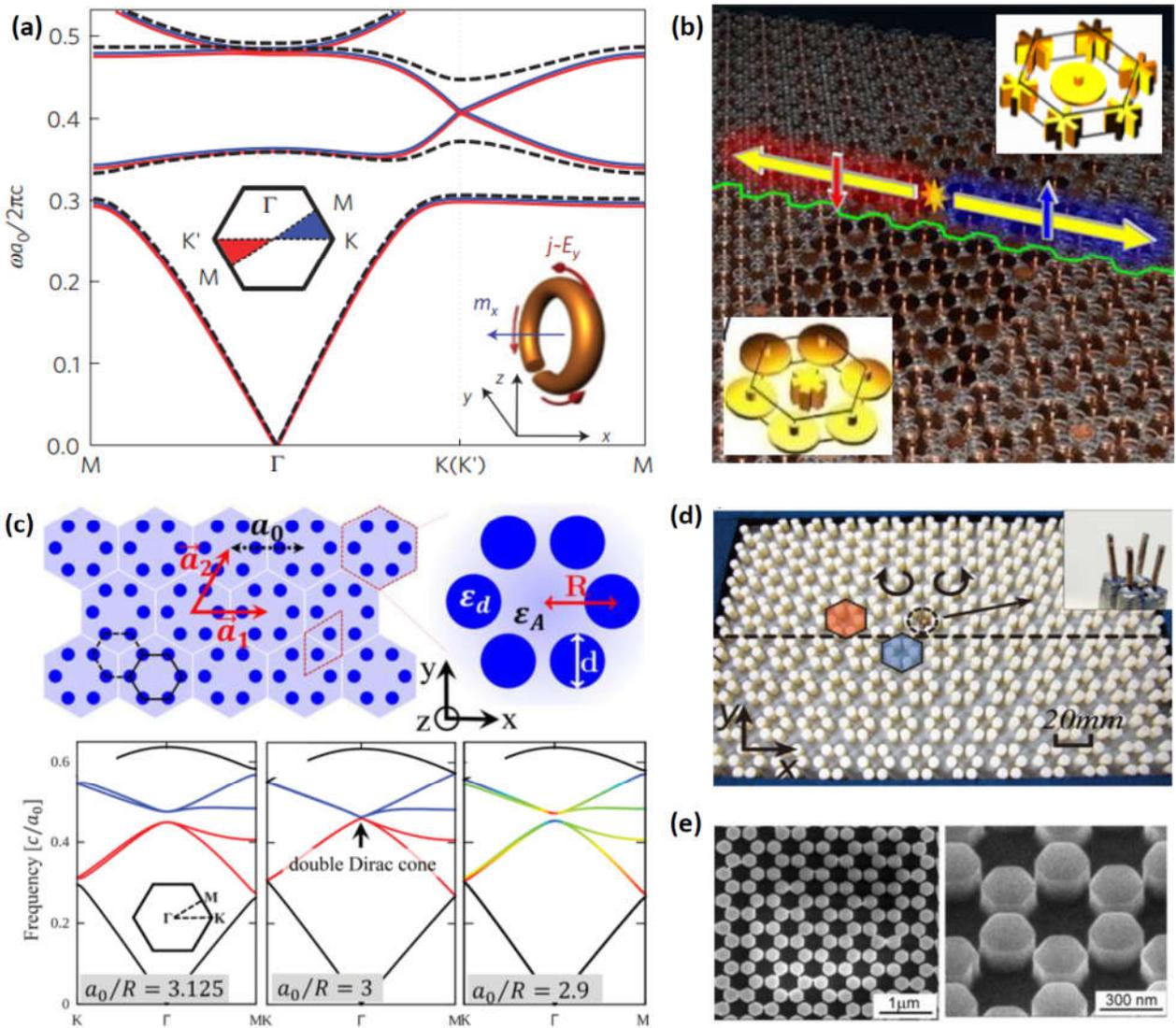

**Figure 4. Quantum spin Hall PCs.** (a) Bulk bands of the quantum spin Hall PC realizing by a triangular bianisotropic PC [119]. (b) Experimental realization of the quantum spin Hall PC in a metacrystal waveguide and the unidirectional transport of spin-polarized edge states at the boundary [120]. (c) Realization of quantum spin Hall PC in an all-dielectric PC whose unit cell contains six cylinders [130]. Bulk bands show the topological phase transition which is realized by changing the distance between neighboring cylinders. (d, e) The experimental realization of the quantum spin Hall PC operating at (d) the microwave spectral range [131] and (e) the visible spectral range [133].

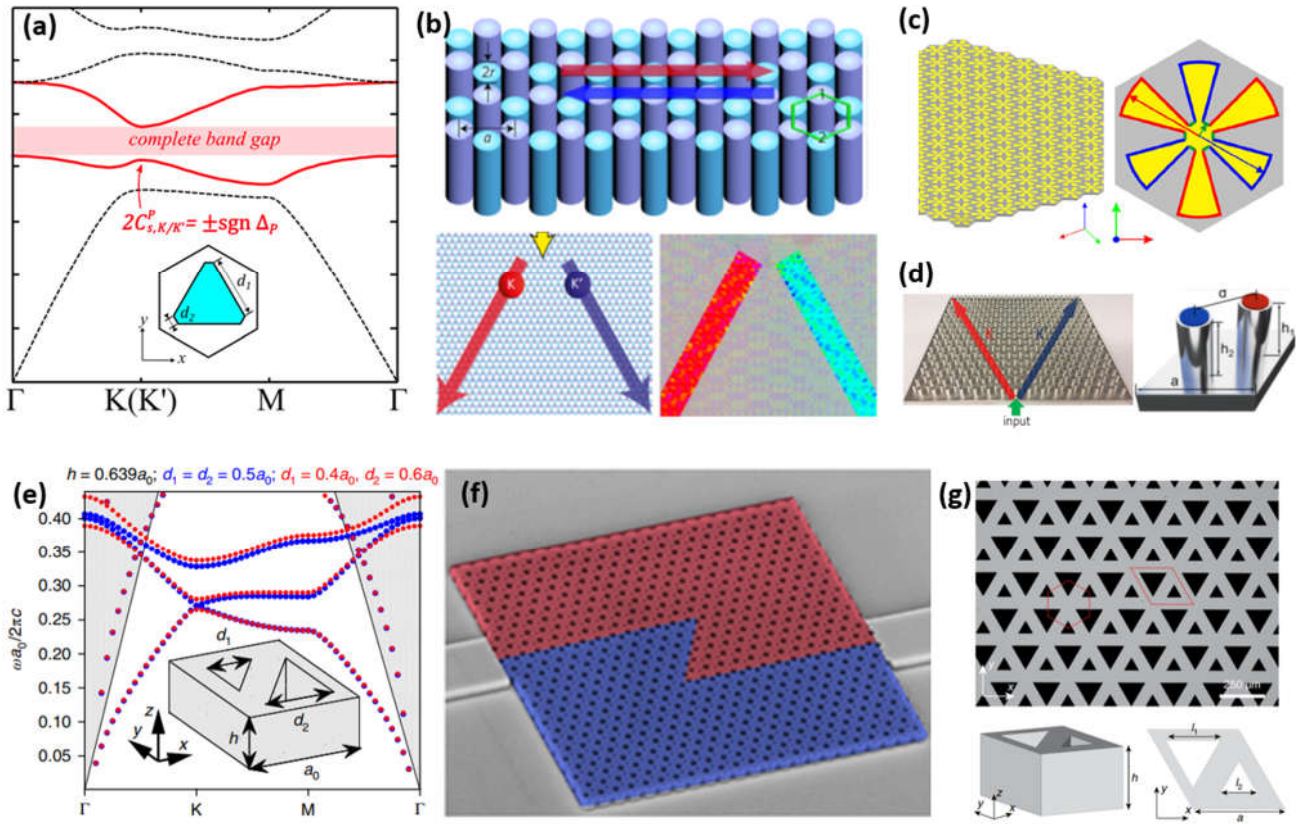

**Figure 5. Quantum valley Hall PCs.** (a) Proposal of quantum valley Hall PCs in the 2D dielectric PCs [143]. Bulk bands with nonzero valley Chern number are obtained by breaking the mirror or inversion symmetry. (b) Valley-spin locking behavior in VPCs with bianisotropy [144] (c-g) Realization of quantum VPCs in the (c, d) microwave frequency [27, 145], (e, f) near-infrared frequency [158, 159], and (g) terahertz frequency [160].

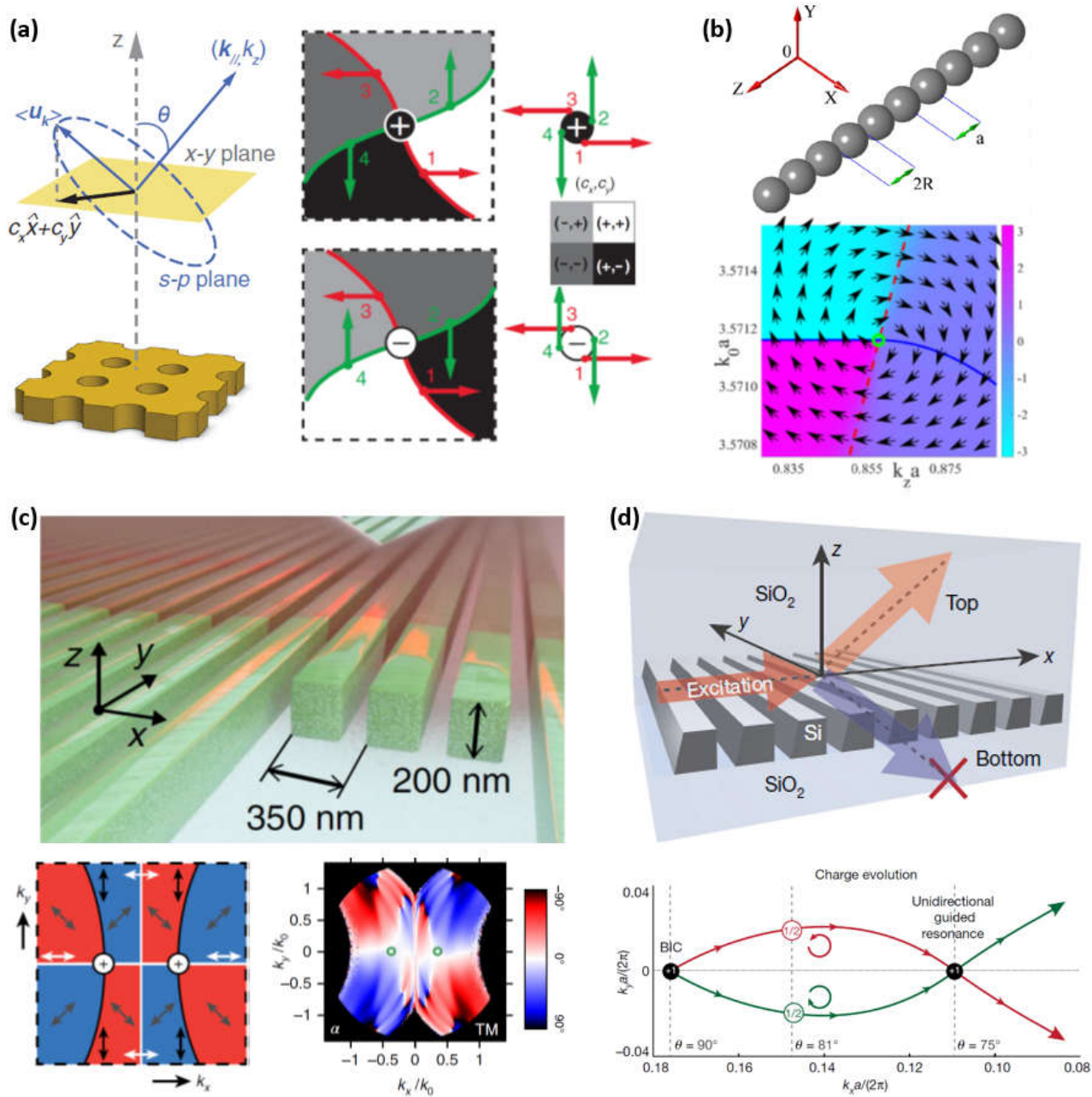

**Figure 6. BICs in PCs.** (a) Characterization of topological charge of BICs in PCs [170]. (b) Periodic array of dielectric spheres and the representation of BIC in the k space [232]. (c) Top: design of the 1D $Si_3N_4$ grating. Bottom: sketch of the expected polarization of the TM-mode radiation (left) and corresponding experimental results (right) in wave vector space [217]. (d) Top: topologically enabled unidirectional guided resonances in 1D grating. Bottom: charge evolution from BIC to unidirectional guided resonance [218].

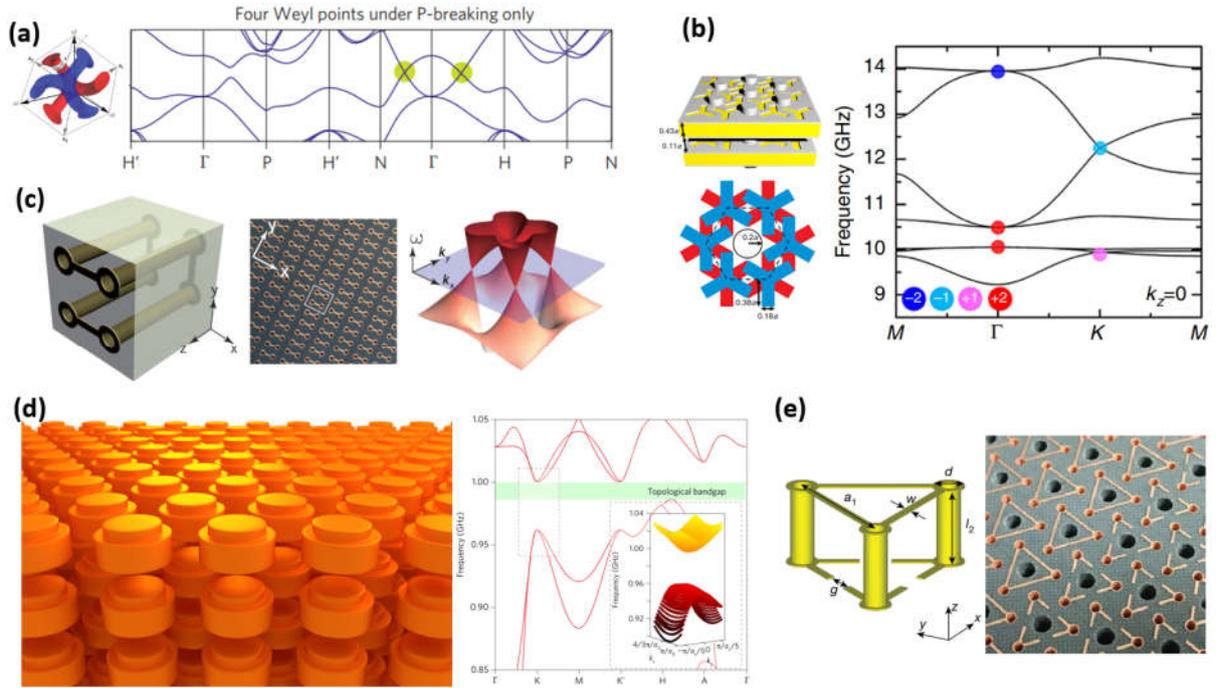

**Figure 7. Weyl points and topological bandgap in 3D PCs.** (a) Frequency isolated Weyl points are obtained in the 3D gyroid PCs [237]. (b) Multi Weyl points are realized in stacked metallic PCBs with interlayer coupling [238]. (c) Ideal Weyl points are achieved in a 3D PC [38]. (d) Theoretical proposal [247] and (e) Experimental realization [248] of weak 3D PTIs with a topological bandgap.

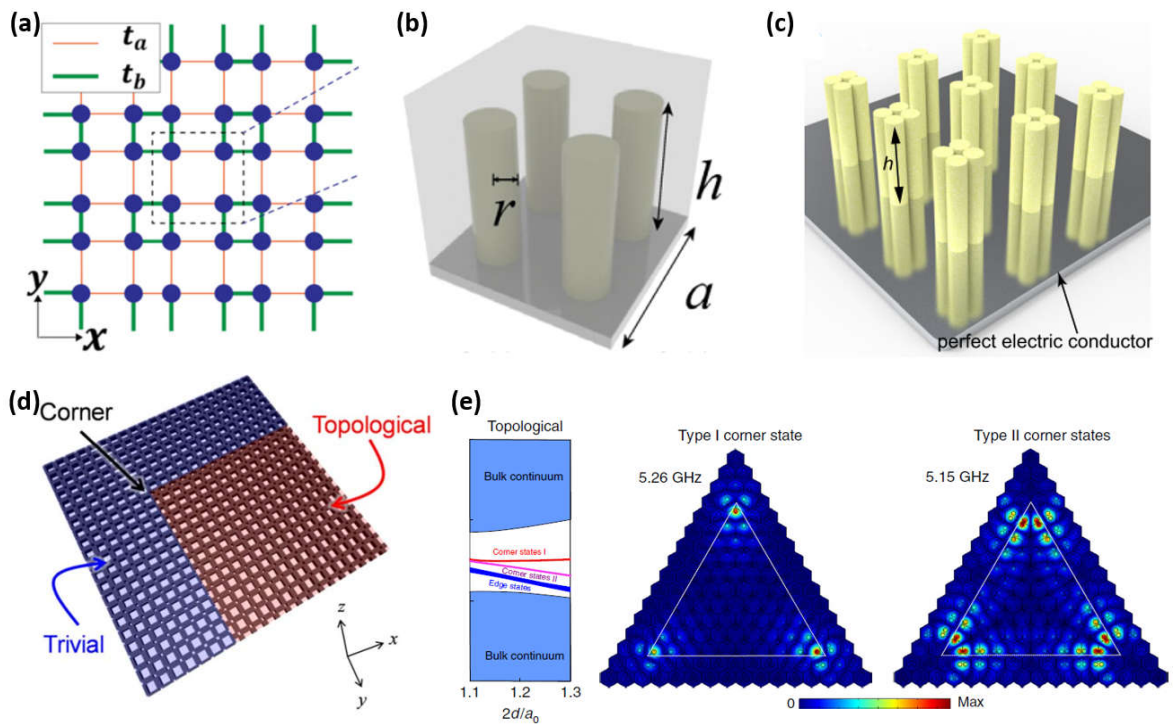

**Figure 8. Higher order PCs.** (a) Proposal of a second-order PC in the 2D PC whose unit cell has four dielectric rods [264]. (b, c) Realization of a second-order PC in the 2D PC [267] and PC slab [268] at the microwave frequency. (d) Realization of a second-order PC whose corner states operates at the near-infrared frequency [266]. (e) Type I and type II corner states of the triangular kagome PC [269].

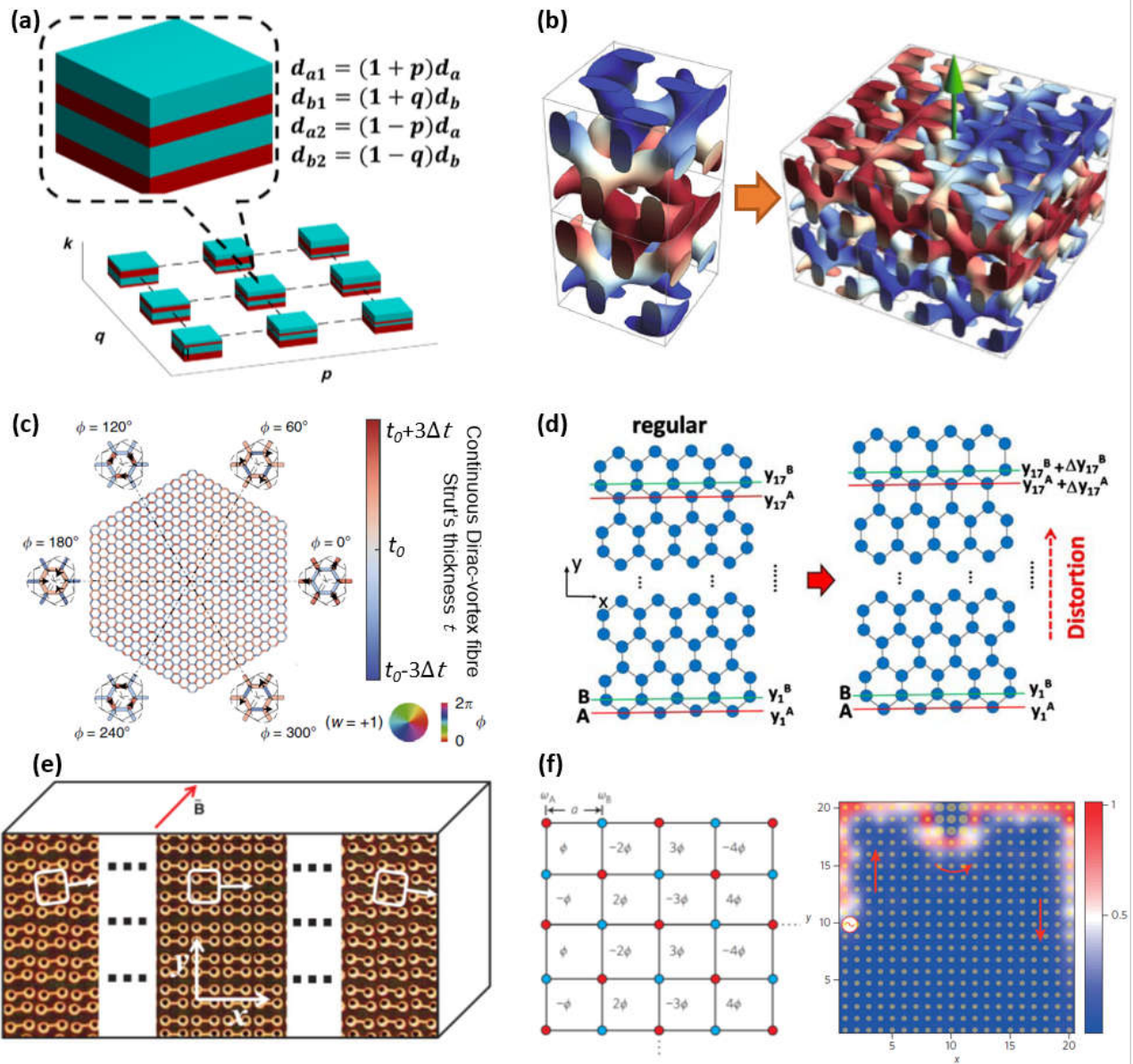

**Figure 9. TPCs with synthetic dimensions, deformations and modulations.** (a) 1D PCs with different $p$ and $q$ values. The combination of $k$, $p$ and $q$ forms a 3D parameter space where synthetic Weyl points are constructed [79]. (b) Left: the double gyroid PC with plain modulated volume fraction (blue and red). Right: the double gyroid structure with single helix modulation, whose helix center supports an one-way fiber mode [245]. (c) The structure of a Dirac-vortex PC fiber. The color of each strut stands for its width [297]. (d) The honeycomb PC without and with distortion which induces a gauge field [301]. (e) The inhomogeneous 3D Weyl metamaterials supporting chiral zero modes. The gradient of metallic coils' rotation angle in the $x$ direction induces a gauge field. The corresponding artificial magnetic field is in the $z$ direction [39]. (f) The photonic resonator lattice exhibiting an effective magnetic field generated with temporal modulation, and featuring with an one-way edge mode [75].

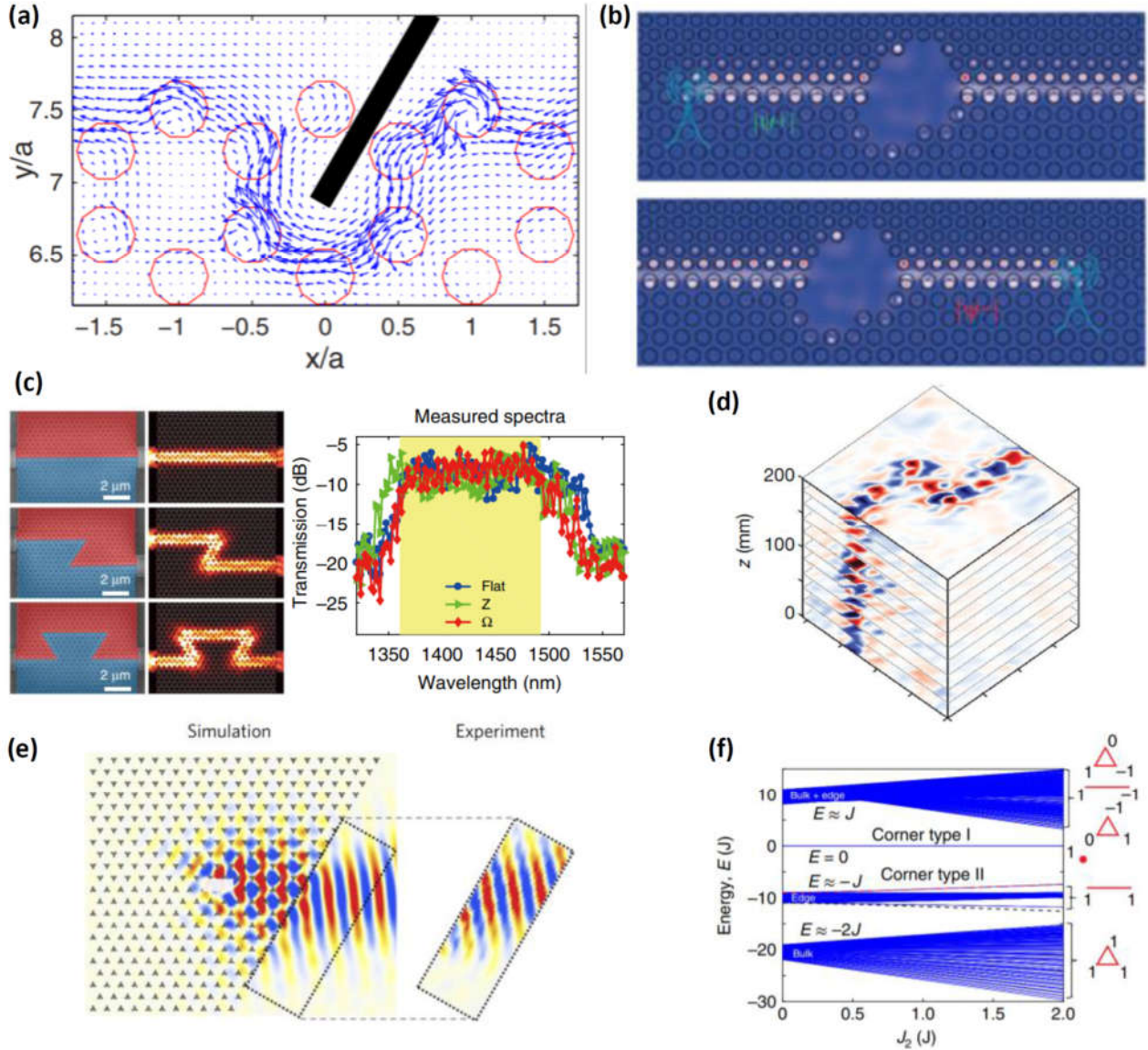

**Figure 10. Robustness of TPCs.** (a) One-way propagation of chiral edge states which are insensitive to imperfections on the zigzag edge of the honeycomb magnetic PC [109]. (b) One-way propagation of the spin-polarized helical edge states which propagate through the cavity obstacle [119]. (c) Robust transport of valley-dependent edge states which are backscattering immune in waveguides with sharp bends [159]. (d) Robust propagation of topological surface states along a non-planar surface in 3D PTIs [248]. (e) Topologically protected refraction of kink states in VPCs [140]. (f) Topological type-I corner states are pinned to zero energy in the second-order PCs [269].

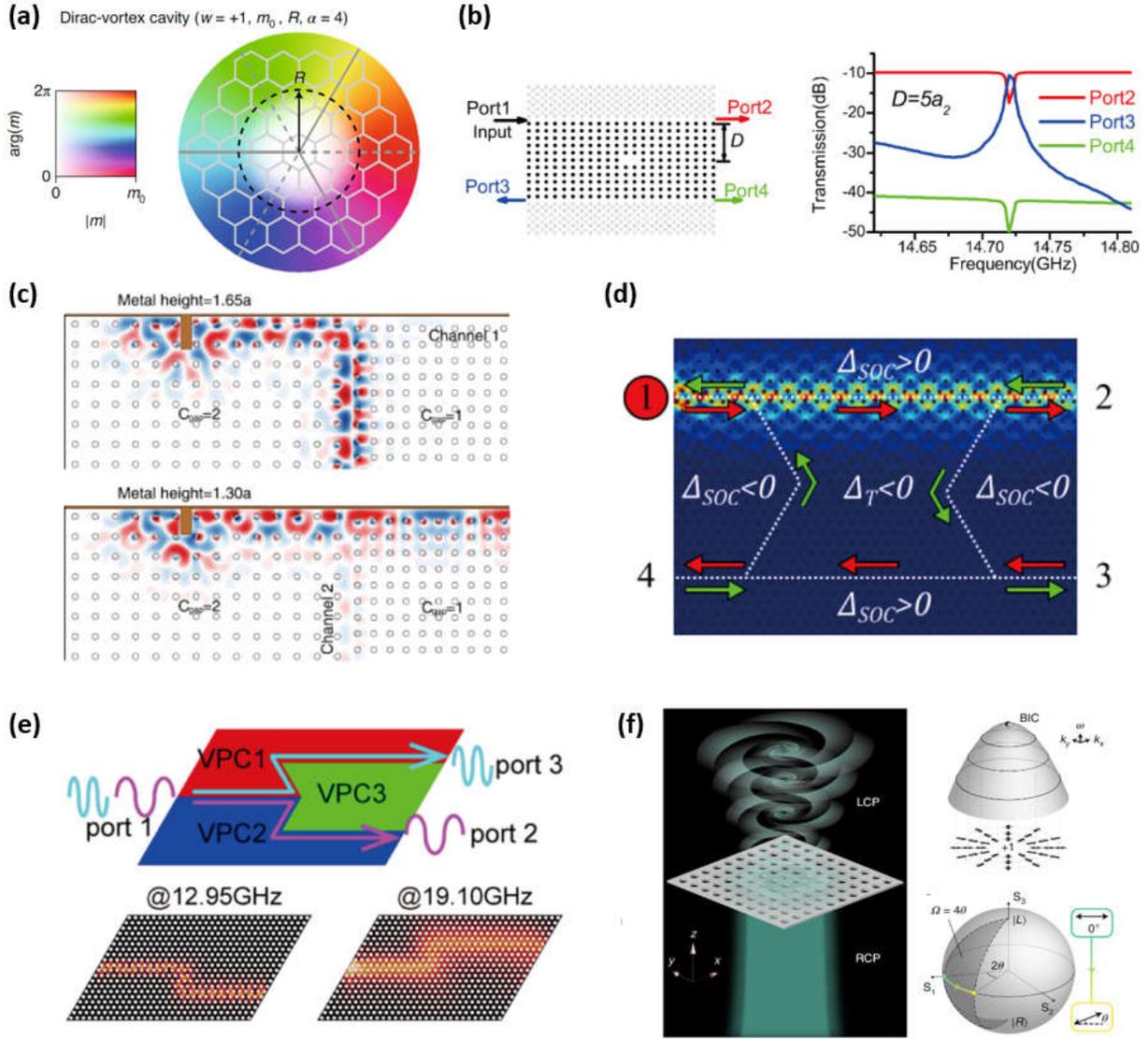

**Figure 11. Prototype of photonic devices made by TPCs.** (a) Dirac vortex cavity with Kekulé modulation characterized by complex value **m** [300]. (b) Schematic and the transmittance of the unidirectional channel-drop four-port filter by one-way magnetic PC waveguides [115]. (c) A power splitter implemented with $C_{gap} = 2$ and $C_{gap} = 1$ magnetic PCs bordered on the top by a metallic wall [111]. (d) A four-port broadband circulator based on nonreciprocal topologically protected edge waves between QSH-PTIs and a QH-PTI [138]. (e) A wavelength demultiplexer based on VPCs with dual bandgaps [327]. (f) Optical vortex generator constructed with PC slab supporting BICs [192].

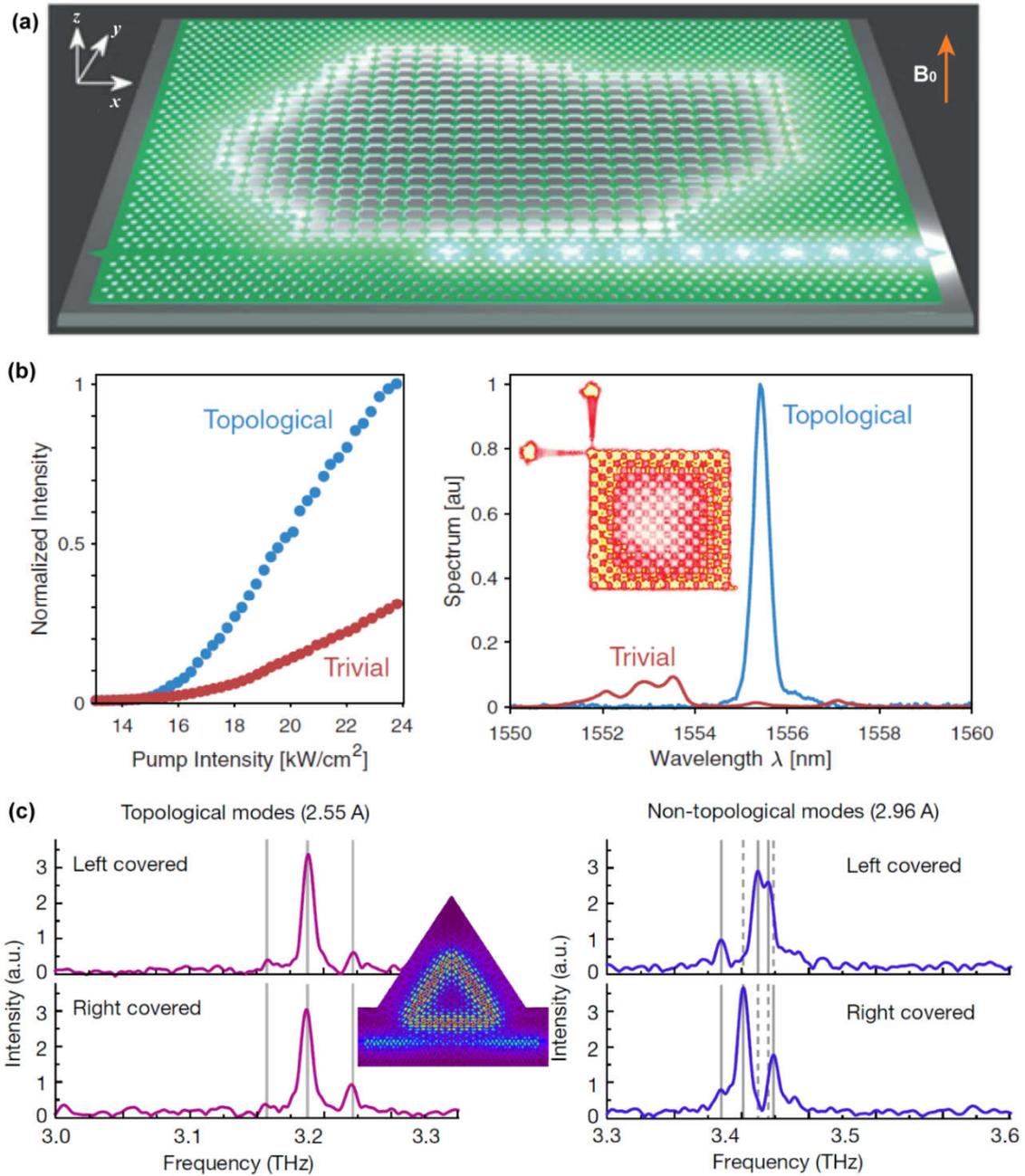

**Figure 12. Lasing in cavities based on 1D states in topological photonic systems.** (a) Schematic of nonreciprocal lasing in an arbitrarily shaped and integrated topological cavity in which time-reversal symmetry is broken by applying an external magnetic field. The InGaAsP multiple quantum wells are bonded on yttrium iron garnet that had been grown on gadolinium gallium garnet [328]. (b) Topological laser in coupled ring resonators on an InGaAsP quantum wells platform. The topological structure has higher output intensity and narrow linewidth compared with trivial counterpart. Inset: image of the lasing pattern (topological edge mode) in an array of topologically coupled resonators and the output ports [337]. (c) Emission spectra for topological laser with valley edge modes in a directional outcoupling configuration [331]. The peak intensities of the left and right covered topological lasing modes are similar while the spectra of the non-topological modes are completely different.

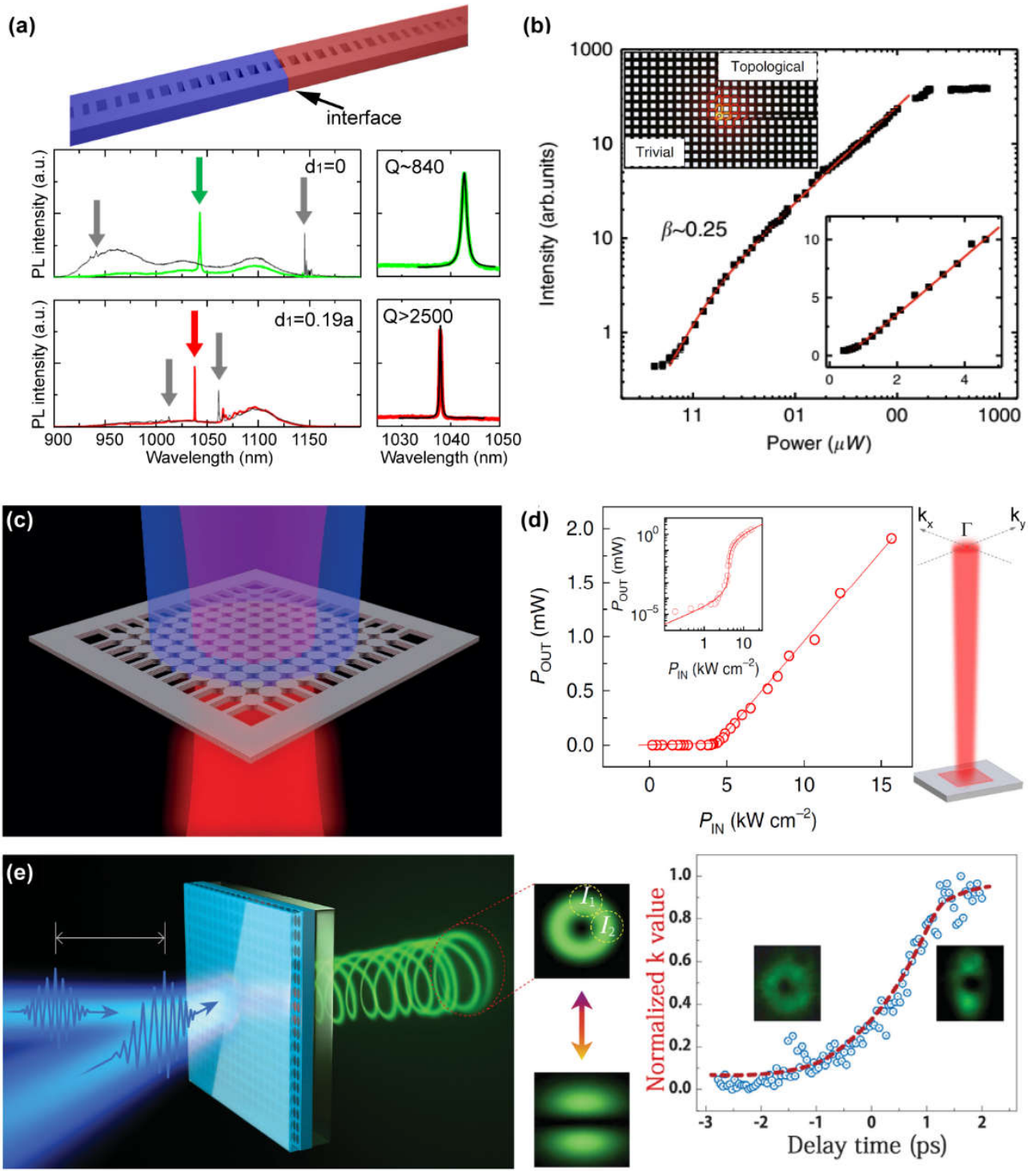

**Figure 13. Lasing in nanocavities based on 0D states and vertical emission in TPCs.** (a) Schematic of 1D TPC nanocavity (top), and measured photoluminescence (PL) spectra and Q factors for the nanobeams in TPCs with different parameters (middle and bottom) [334]. (b) The pump-power dependence of nanolasers based on the second-order corner states. The insets show the field pattern of corner states and the enlarged curve around the threshold of lasing [332]. (c) Lasing action from BICs in PC slab [176]. (d) Topological bulk laser based on the band-inversion-induced reflection [330]. (e) Ultrafast controlling of quasi-BIC vortex microlasers with the delay time of two incident pulses [191].

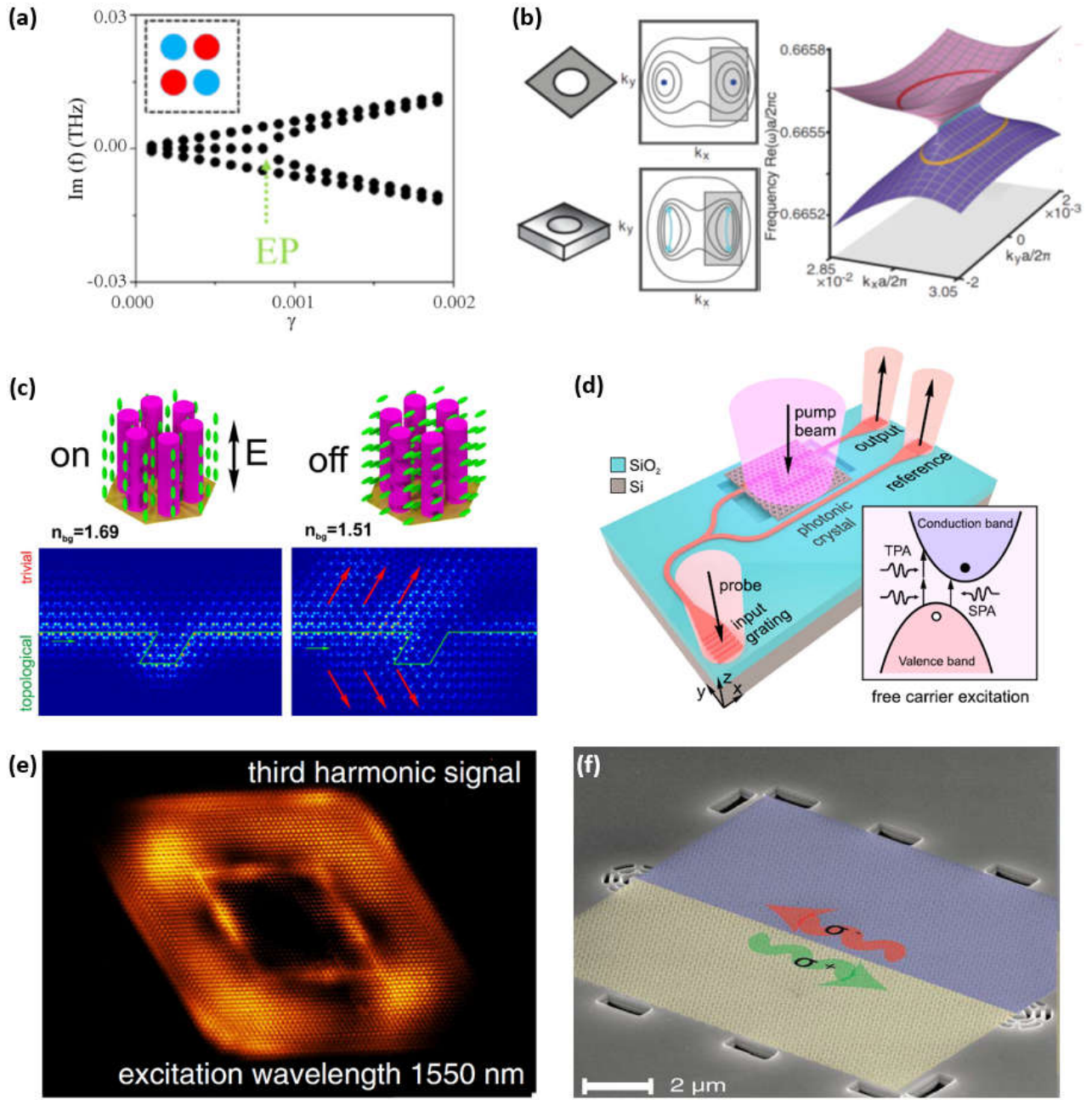

**Figure 14. Active applications of TPCs.** (a) The dependence between the imaginary part of eigenfrequencies and the gain/loss parameter γ in a square PC whose unit cell has four rods with gain or loss [354]. The exceptional point pointed by a green arrow indicates the position where two eigenmodes become nondegenerate. (b) Bulk Fermi arc arising from paired exceptional points splits from a single Dirac point [355]. (c) An optical switch based on reconfigurable TPCs. The optical switch is controlled by applying voltage to the electrodes [360]. (d) Tuning of transmission of the topological PC is enabled by refractive index modulation due to optically induced free carrier excitation [362]. (e) Third-harmonic intensity distribution image in the all-dielectric TPC [369]. (f) Scanning electron microscope image of the topological quantum optics interface. The supported helical edge states with opposite circular polarization has different transport directions [136].